\documentclass[twocolumn]{aastex62}

\graphicspath{{./}{figures/}}
\usepackage{float}
\usepackage[caption = false]{subfig}
\usepackage[]{threeparttable}
\usepackage{lineno}
\usepackage{appendix}
\newcommand{\hi}{H\,{\sc i}}

\shortauthors{Mutlu-Pakdil et al.}
\begin{document}

\title{The Faint Satellite System of NGC~253: Insights into Low-Density Environments and No Satellite Plane\footnote{This paper includes data gathered with the 6.5~m Magellan Telescope at Las Campanas Observatory, Chile.}}

\author[0000-0001-9649-4815]{Bur\c{c}in Mutlu-Pakdil}
\affil{Department of Physics and Astronomy, Dartmouth College, Hanover, NH 03755, USA}

\author[0000-0003-4102-380X]{David J. Sand}
\affil{Department of Astronomy/Steward Observatory, 933 North Cherry Avenue, Rm. N204, Tucson, AZ 85721-0065, USA}

\author[0000-0002-1763-4128]{Denija Crnojevi\'{c}}
\affil{University of Tampa, 401 West Kennedy Boulevard, Tampa, FL 33606, USA}

\author[0000-0001-8354-7279]{Paul Bennet}
\affiliation{Space Telescope Science Institute, 3700 San Martin Drive, Baltimore, MD 21218, USA}

\author[0000-0002-5434-4904]{Michael G. Jones}
\affiliation{Steward Observatory, University of Arizona, 933 North Cherry Avenue, Rm. N204, Tucson, AZ 85721-0065, USA}

\author[0000-0002-0956-7949]{Kristine Spekkens}
\affil{Department of Physics and Space Science, Royal Military College of Canada P.O. Box 17000, Station Forces Kingston, ON K7K 7B4, Canada}
\affil{Department of Physics, Engineering Physics and Astronomy, Queen's University, Kingston, ON K7L 3N6, Canada}

\author[0000-0001-8855-3635]{Ananthan Karunakaran}
\affiliation{Department of Astronomy \& Astrophysics, University of Toronto, Toronto, ON M5S 3H4, Canada}
\affiliation{Dunlap Institute for Astronomy and Astrophysics, University of Toronto, Toronto ON, M5S 3H4, Canada}

\author[0000-0002-5177-727X]{Dennis Zaritsky}
\affil{Department of Astronomy/Steward Observatory, 933 North Cherry Avenue, Rm. N204, Tucson, AZ 85721-0065, USA}

\author[0000-0003-2352-3202]{Nelson Caldwell}
\affil{Center for Astrophysics, Harvard \& Smithsonian, 60 Garden Street, Cambridge, MA 02138, USA}

\author[0000-0001-8245-779X]{Catherine E. Fielder}
\affiliation{Steward Observatory, University of Arizona, 933 North Cherry Avenue, Rm. N204, Tucson, AZ 85721-0065, USA}

\author[0000-0001-8867-4234]{Puragra Guhathakurta}
\affiliation{Department of Astronomy and Astrophysics, University of California Santa Cruz, University of California Observatories, 1156 High Street, Santa Cruz, CA 95064, USA}

\author[0000-0003-0248-5470]{Anil C. Seth}
\affil{University of Utah, 115 South 1400 East Salt Lake City, UT 84112-0830, USA}

\author[0000-0002-4733-4994]{Joshua D. Simon}
\affiliation{Observatories of the Carnegie Institution for Science, 813 Santa Barbara Street, Pasadena, CA 91101, USA}

\author[0000-0002-1468-9668]{Jay Strader}
\affil{Department of Physics and Astronomy, Michigan State University,East Lansing, MI 48824, USA}

\author[0000-0001-6443-5570]{Elisa Toloba}
\affil{Department of Physics, University of the Pacific, 3601 Pacific Avenue, Stockton, CA 95211, USA}

\email{Burcin.Mutlu-Pakdil@dartmouth.edu}

\begin{abstract}

We have conducted a systematic search around the Milky Way (MW) analog NGC~253 (D=3.5~Mpc), as a part of the Panoramic Imaging Survey of Centaurus and Sculptor (PISCeS) -- a Magellan$+$Megacam survey to identify dwarfs and other substructures in resolved stellar light around MW-mass galaxies outside of the Local Group. In total, NGC~253 has five satellites identified by PISCeS within 100~kpc with an absolute V-band magnitude $M_V<-7$. We have additionally obtained deep Hubble Space Telescope imaging of four reported candidates beyond the survey footprint: Do~III, Do~IV, and dw0036m2828 are confirmed to be satellites of NGC~253, while SculptorSR is found to be a background galaxy. We find no convincing evidence for the presence of a plane of satellites surrounding NGC~253. We construct its satellite luminosity function, which is complete down to $M_V$$\lesssim$$-8$ out to 100~kpc and $M_V$$\lesssim$$-9$ out to 300~kpc, and compare it to those calculated for other Local Volume galaxies. Exploring trends in satellite counts and star-forming fractions among satellite systems, we find relationships with host stellar mass, environment, and morphology, pointing to a complex picture of satellite formation, and a successful model has to reproduce all of these trends. 

\end{abstract}


\section{Introduction} \label{sec:intro}

Dwarf galaxies are unique laboratories to study the physics of dark matter and galaxy evolution. The currently favored $\Lambda$ Cold Dark Matter ($\Lambda$CDM) cosmological model predicts the existence of a hierarchy of dark matter halos, within which galaxies form and reside \citep[e.g.,][]{Wechsler2018}. This theory is strongly supported by observations at large spatial scales, but there are important open questions at small scales, corresponding to that of dwarf galaxies \citep[see][]{Sales2022}. Because of the detail with which they can be studied, Local Group satellites have been the primary sample for understanding the astrophysics and cosmological implications of dwarf galaxies (\citealt{Nadler2021}, among others). However, there is a danger of ‘over-tailoring’ the models to fit local observations. To fully test the $\Lambda$CDM model and its underlying astrophysics (e.g., stellar and supernova feedback, reionization, tidal and ram pressure stripping, etc.), studies of satellite systems beyond the Local Group are necessary to sample primary halos with a range of masses, morphologies, and environments.

Observationally, this work is already underway employing diverse approaches: \hi~observations \citep[e.g.,][]{Cannon2011,Papastergis2015,Yaryura2016}, large-area integrated light surveys around nearby galaxies \citep[e.g.,][]{Muller2015,Danieli17,Bennet17,Park2017,Park2019,Bennet19,Bennet20,Davis21,Carlsten2022,Crosby2023,Fan2023}, focused deep-imaging surveys that allow satellites to be resolved into stars \citep[e.g.,][]{Chiboucas2009,Chiboucas13,Sand14,Sand15b,Crnojevic16,Carlin16,Toloba16,Smercina18,Crnojevic19,carlin21,Drlica-Wagner2021}, and spectroscopic surveys around Milky Way (MW) analogs at larger distances \citep{Geha17,Mao2021}. 

These different approaches complement each other and come with distinct advantages and limitations. Focused deep-imaging surveys facilitate resolved searches for very faint dwarf galaxies ($M_V>-9$) within 4~Mpc, where we can study the effects of reionization and compare our findings with the latest discoveries of the Local Group. It is also possible to get a deeper understanding of their nature from resolved stellar populations. On the other hand, integrated light surveys are very successful in identifying brighter dwarfs ($M_V$$\lesssim$$-9$) within the $\sim$4-10~Mpc range. At these greater distances, however, detailed investigations with resolved stars are no longer feasible. Resolved star studies also have fewer observational biases, as the distance to the candidates can be derived directly via the tip of the red giant branch stars (TRGB, e.g., \citealt{Sand14,Crnojevic16,Toloba16}). While surface brightness fluctuation (SBF) offers an efficient way to get distances to quenched dwarf galaxies in integrated light surveys, it is well known that the SBF technique is not ideal for gas-rich, star-forming systems because their star-forming regions can significantly affect the SBF measurements \citep{Greco2021}. While spectroscopic surveys can probe even greater distances ($\sim25-40$~Mpc), they may have their own biases, potentially favoring objects with emission lines, or disfavoring low surface brightness objects that are difficult to spectroscopically confirm. 

In \citet{MutluPakdil22}, we presented the focused Magellan$+$Megacam deep-imaging survey of the nearest MW analog in a very low-density environment: the edge-on spiral NGC~253 (D = 3.5~Mpc; \citealt{Radburn-Smith2011}, a total stellar mass of $\approx4.4\times10^{10} M_{\odot}$; \citealt{Bailin2011}). While it is usually classified as a member of the Sculptor group, this ``group" is not truly a bound system, but instead, a filament extended along our line-of-sight \citep{Jerjen1998,Karachentsev2003}. Hence NGC~253 is evolving essentially in isolation, providing us a unique opportunity to extend the range of environments probed by the existing surveys in resolved stars. In this previous work, we reported the discovery of three new ultra-faint dwarf galaxies (UFDs, $M_V \gtrsim -7.7$) via a visual search, and presented Hubble Space Telescope ({\it HST}) follow-up observations for these three systems, as well as for two other dwarfs identified in the early stages of our survey \citep{Sand14,Toloba16}. 

A distinctive feature characterizing the population of dwarf galaxies is the luminosity function (LF), i.e., the total number of dwarfs as a function of luminosity. The relationship between the LF and the mass function of dark matter halos probes the physics of galaxy formation in the smallest halos and constrains dark matter models. The observed LF accurately represents the true LF only if the dwarf galaxy sample is complete over the considered luminosity range. However, the detection of faint galaxies varies significantly based on their luminosity, surface brightness, and distance, and many of these galaxies are near the detection limits of the surveys in which they are discovered. Therefore, deriving the LF requires accurately quantifying the sensitivity of dwarf galaxy searches. Similarly, confirming the nature of candidate dwarfs and firmly establishing their membership with the host galaxy are equally vital for constructing an accurate LF. In this paper, our main focus is on the NGC~253 LF and its satellite system as a whole. 

First, in Section~\ref{sec:pisces}, we present the general overview of our survey of NGC~253. In Section~\ref{sec:search}, we present the details of our resolved dwarf search. In Section~\ref{sec:completeness}, we statistically characterize our overall satellite detection efficiency. In Section~\ref{sec:beyond}, we present {\it HST} follow-up of dwarf candidates beyond the survey footprint. In Section~\ref{sec:plane}, we revisit the recently proposed satellite plane around NGC~253. In Section~\ref{sec:LF}, we derive the NGC~253 LF to compare with those calculated for other Local Volume galaxies. In Section~\ref{sec:discussion}, we explore the characteristics and trends among the satellite systems, as a function of the most dominant mergers, host stellar mass, local density environment, and morphology. Finally, we summarize our key results in Section~\ref{sec:conclusion}.

\section{The PISCES Survey} \label{sec:pisces}

The Panoramic Imaging Survey of Centaurus and Sculptor (PISCeS) is a Magellan$+$Megacam survey to search for dwarf galaxies and signs of hierarchical structure formation in the halos of NGC~253 and Centaurus~A (Cen~A) -- two nearby galaxies of different morphologies in two environments substantially different from the Local Group. NGC~253 is a starbursting spiral in a low-density environment ($D$$=$$3.5$~Mpc; \citealt{Radburn-Smith2011}) and Cen~A is an elliptical in a relatively rich group ($D$$=$$3.8$~Mpc; \citealt{Harris2010}). While the PISCeS Cen~A campaign resulted in the discovery of 11 new satellites and several previously unknown streams and shells \citep{Crnojevic16,Crnojevic19}, the NGC~253 campaign led to the discovery of 5 new satellites  \citep{Sand14,Toloba16,MutluPakdil22}. Our previous papers on the NGC~253 PISCeS data were dedicated to these dwarf discoveries, providing a detailed analysis of their stellar populations and physical properties. In this paper, our main goal is to statistically characterize the overall satellite detection efficiency in our PISCeS NGC~253 footprint, derive the satellite luminosity function of NGC~253, and place its satellite system in the galaxy formation context. 

\begin{figure*}
\centering
\includegraphics[width = 0.80\linewidth]{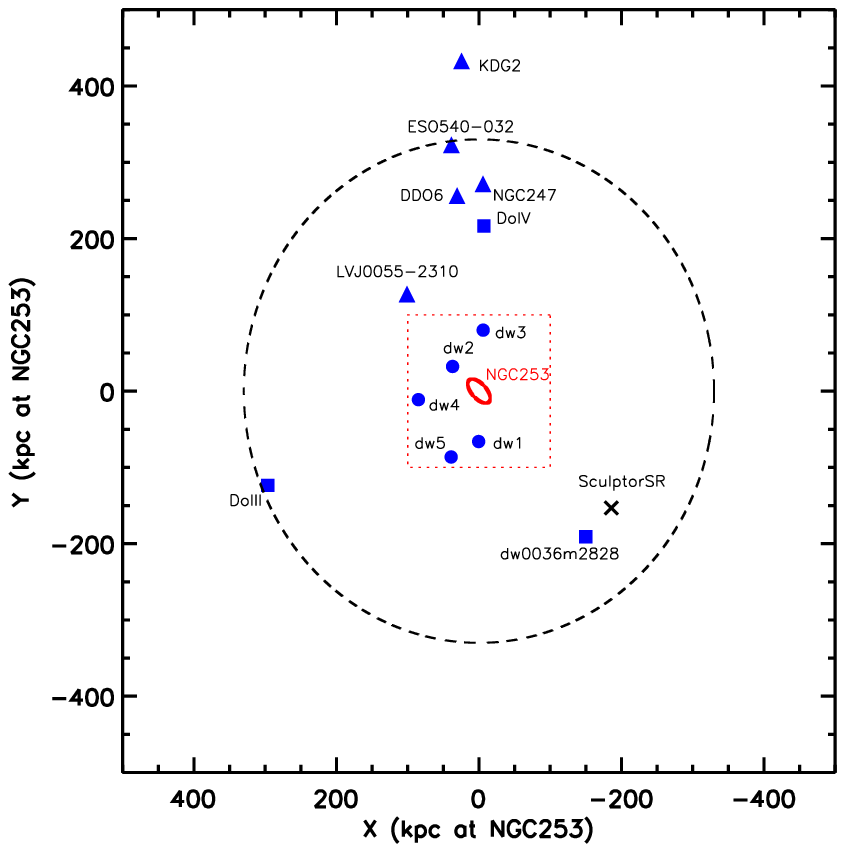}
\caption{Footprint of PISCeS (red dashed line), centered on NGC~253 (red ellipse). Confirmed dwarfs at the distance of NGC~253 are shown with filled blue symbols, five were discovered in PISCeS (circles): Scl-MM-dw1, Scl-MM-dw2, Scl-MM-dw3, Scl-MM-dw4, and Scl-MM-dw5 \citep{Sand14,Toloba16,MutluPakdil22}. The blue squares are the three dwarfs recently discovered beyond PISCeS and confirmed as NGC~253 satellites by our {\it HST} follow-up observations: Do~III, Do~IV, and dw0036m2828 (\citealt{Martinez-Delgado21,Carlsten2022}, see Section~\ref{sec:beyond}). The remaining five (triangles) are previously known dwarfs from \citet{Karachentsev2021}. The black cross ($\times$) represents SculptorSR which turns out to be a background object based on our {\it HST} imaging (see Appendix). The black circle represents the approximate virial radius of NGC~253 (330~kpc, \citealt{Mutlupakdil21}).
\label{fig:footprint}}
\end{figure*}

As part of PISCeS, we have observed 81 Megacam fields around NGC~253, which reach out to a projected radius of $\sim$$100$~kpc ($\sim$$1/3$ of its virial radius, \citealt{Mutlupakdil21}; see Figure~\ref{fig:footprint} for the survey footprint). Megacam has a $\sim24\arcmin\times24\arcmin$ field of view (FoV) and a binned pixel scale of 0.16\arcsec. PISCeS typically observes each field for $6\times300$~s in each of the $g$ and $r$ bands to achieve image depths of $g,r\approx26.5$~mag, which is $\sim$2 magnitudes below the TRGB at the distance of NGC~253. Throughout the survey, the median seeing is $\sim$$0.8\arcsec$ in both bands, with the best/worst seeing being $\sim$0.5\arcsec$/1.3\arcsec$ in both bands. The data are reduced in a standard way by the Smithsonian Astrophysical Observatory Telescope Data Center (see \citealt{McLeod15,Crnojevic16}, for further details).

We perform point spread function (PSF)-fitting photometry on each of the stacked final images, using the suite of programs DAOPHOT and ALLFRAME \citep{Stetson87,Stetson94}, following the same methodology described in \citet{Crnojevic16}. We remove objects that are not point sources by culling our ALLFRAME catalogs of outliers in $\chi$ versus magnitude, magnitude error versus magnitude, and sharpness versus magnitude space.  Instrumental magnitudes are then calibrated by matching them to the DES DR2 catalog \citep{DESDR2}. The final calibrated catalogs are dereddened on a star-by-star basis using the \citet{Schlegel98} reddening maps with the coefficients from \citet{Schlafly11}. The extinction-corrected photometry is used throughout this work. 

To assess the photometric uncertainties and incompleteness of our wide-field dataset, we run a series of artificial star tests with the DAOPHOT routine ADDSTAR. We place artificial stars into our images on a regular grid ($\sim$10--20 times the image FWHM). We assign the $r$ magnitude of the artificial stars randomly from 18 to 29 mag with an exponentially increasing probability toward fainter magnitudes, and the $g$ magnitude is then randomly selected with uniform probability based on the $g-r$ color over the range -0.5--1.5~mag. Ten iterations are performed on each field for a total of $\sim$100,000 artificial stars each. Their photometry is derived exactly in the same way as for the real data, and the same quality cuts and calibration are applied. The 50\% completeness limit per pointing varies from $r$$\sim$23.9--27.4~mag and $g$$\sim$25.9--27.8~mag, with the average at $r$$\sim$26.7~mag and $g$$\sim$27.2~mag. 

As our focus in this paper is solely on satellite dwarfs, we will present the global color-magnitude diagram (CMD) and NGC~253 resolved stellar halo properties in a future paper.

\section{Dwarf Satellite Search}\label{sec:search}
\begin{figure*}[!ht]
\centering
{\includegraphics[width = 2.25in]{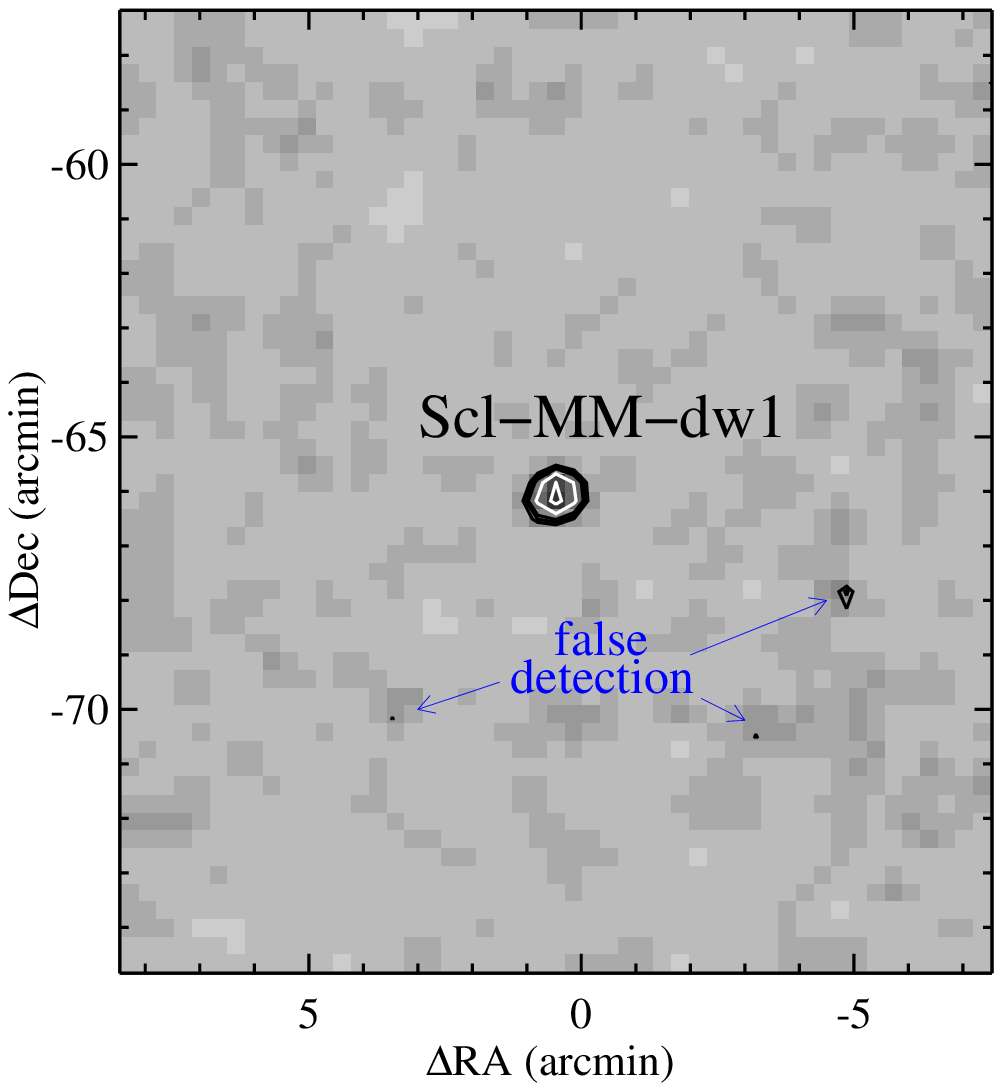}}
{\includegraphics[width = 2.25in]{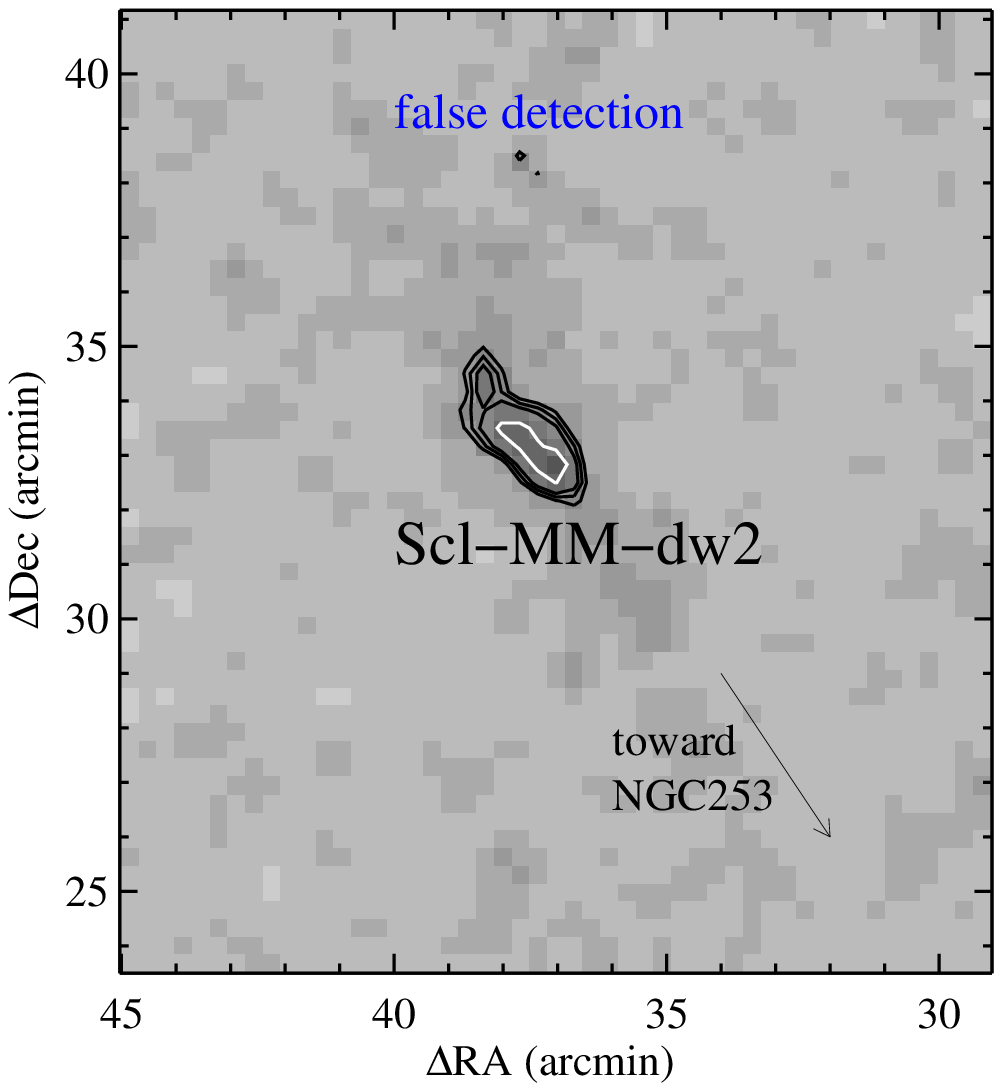}}
{\includegraphics[width = 2.25in]{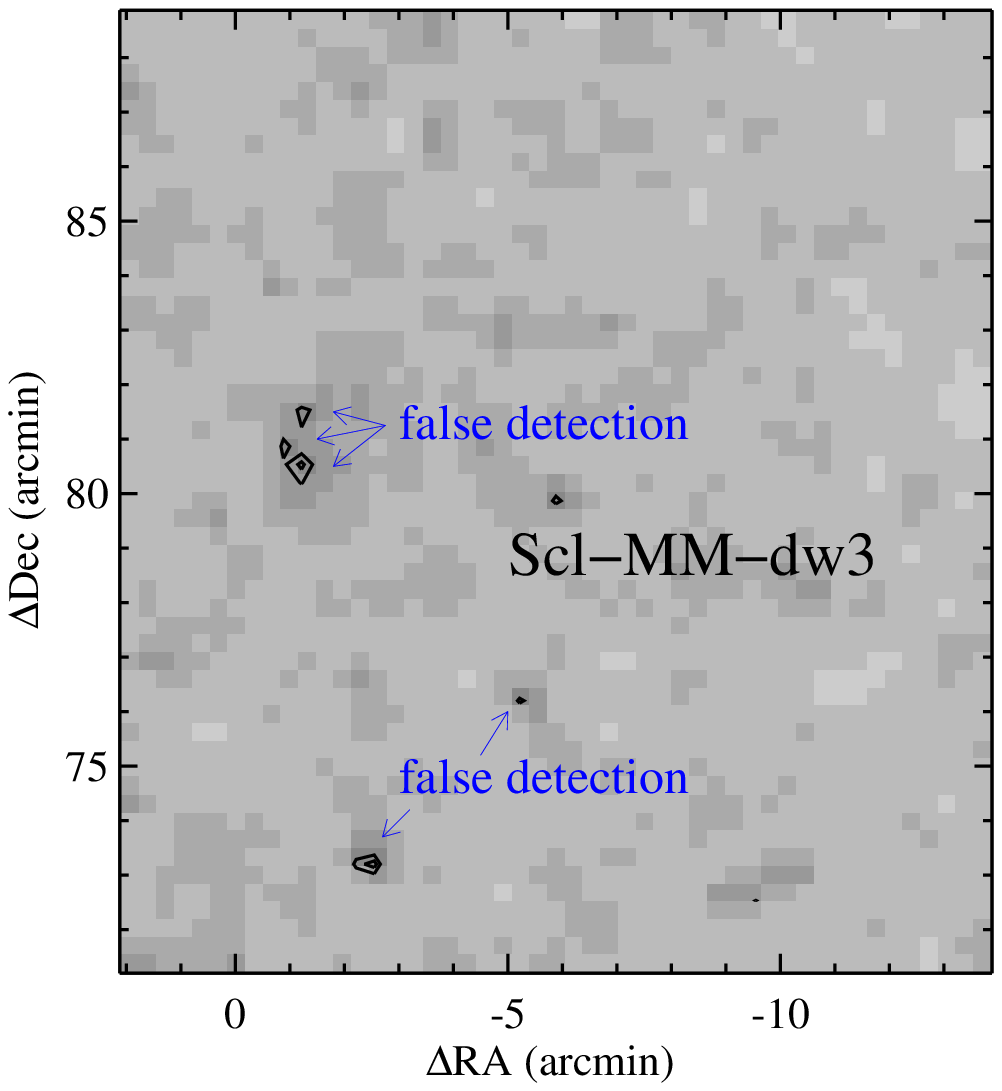}}
\caption{The smoothed matched-filter stellar density maps where the overdensities of previously reported PISCeS dwarfs are recovered (left: Scl-MM-dw1, middle: Scl-MM-dw2, right: Scl-MM-dw3). Positions here are relative to the center of NGC~253. The contour levels show the $5\sigma$, $6\sigma$,$7\sigma$, $10\sigma$, $15\sigma$, and $20\sigma$ levels above the model value. False detections near each system are labeled, and they are mostly shredded bright background galaxies based on visual inspection. We note that Scl-MM-dw1 is detected at $30\sigma$, Scl-MM-dw2 at $17\sigma$, and Scl-MM-dw3 at $5.5\sigma$. \label{fig:detection}}
\end{figure*}

Beyond the Local Group, only the intrinsically brightest stars can be resolved. Given that all known dwarf galaxies contain old ($\gtrsim$10~Gyr) red giant branch (RGB) stars \citep{Weisz2011}, they are the best tracer to use to find these more distant dwarfs. Our PISCeS program is designed to detect satellites as overdensities of RGB stars at the distance of NGC~253. To automatically detect such overdensities and do extensive completeness checks, we adopt a matched-filter technique \citep{Rockosi2002,Walsh2009}, which maximizes the signal to noise in possible dwarf stars over the background. 

Our search algorithm is described in detail in \citet{Mutlupakdil21} and is applied to the PISCeS NGC~253 fields, but we provide a brief overview of the important steps in this section. First, we build well-populated signal CMDs (of $\approx$ 75,000 stars), including the stellar completeness and photometric uncertainties of each field based on artificial star tests, by adopting an old metal-poor Dotter isochrone (i.e., \citealt{Dotter2008}, age$=$ 10~Gyr, [Fe/H]$=-2.0$) and its associated luminosity function. Background CMDs should ideally be chosen from a field far beyond a dwarf galaxy. However, because we do not know which region is free of dwarf galaxy stars, we use all of the stars in each field for background CMDs\footnote{In cases where the contamination from the stellar halo is significant, experimenting with different background fields far from NGC~253 shows no significant changes in our results.} and normalize them based on the area selected. We bin these CMDs into 0.1$\times$0.1 color-magnitude bins. We then spatially bin our stars into 20\arcsec~pixels, smooth our final values using a Gaussian of width of the pixel size, and create our final smoothed matched-filter maps. The MMM routine in IDL is used to calculate the background level (sky$_{mean}$) and variance (sky$_{sigma}$) of these smoothed maps. The normalized signal can be defined as $S = (smooth map  - sky_{mean})/sky_{sigma}$, and gives the number of standard deviations ($\sigma$) above the local mean. We use $S$ as a measure of detection signal and visually inspect any stellar overdensity with $S > 5\sigma$.  

The number of detected sources ranges from 2 to 20 depending on the field. After the visual inspection, we find that the majority of these detections are false positives, primarily bright background galaxies that have been detected as multiple point sources by DAOPHOT. While no new dwarf candidates pass our visual inspection, two previously reported PISCeS classical dwarfs are easily recovered: Scl-MM-dw1 and Scl-MM-dw2 (see Figure~\ref{fig:detection}). The other three reported PISCeS dwarfs are UFDs, and were identified via systematic visual search  \citep{MutluPakdil22}. While Scl-MM-dw3 ($M_V=-7.2$) is detected just above our threshold cut with 5.5$\sigma$, the other two fall in the fields with the worst seeing cases and thus could not be recovered using our matched-filter method. 

While the matched-filter technique is adept at identifying well-resolved systems, its efficacy diminishes when dealing with semi-resolved systems - those characterized by an underlying diffuse light contribution with a few resolved stars overlaid. Although visual searches are more successful in such cases (as in \citealt{MutluPakdil22}), without an automated search algorithm, a statistical sensitivity analysis is not possible. Recently, \citet{Jones2023} introduced a new optimized search algorithm for semi-resolved systems based on a convolutional neural network classifier, which can serve as a powerful and complementary tool for finding and studying these elusive objects.

In the following section, we quantify the detectability of dwarf galaxies as a function of size and luminosity, and we do not include satellites that fall below our completeness limits in our derived satellite LF and related discussions (see Section~\ref{sec:LF}). 

\section{Completeness Tests}\label{sec:completeness}

\begin{figure*}[!t]
\centering
%{\includegraphics[width=7in]{figure3.png}}
{\includegraphics[width=7in]{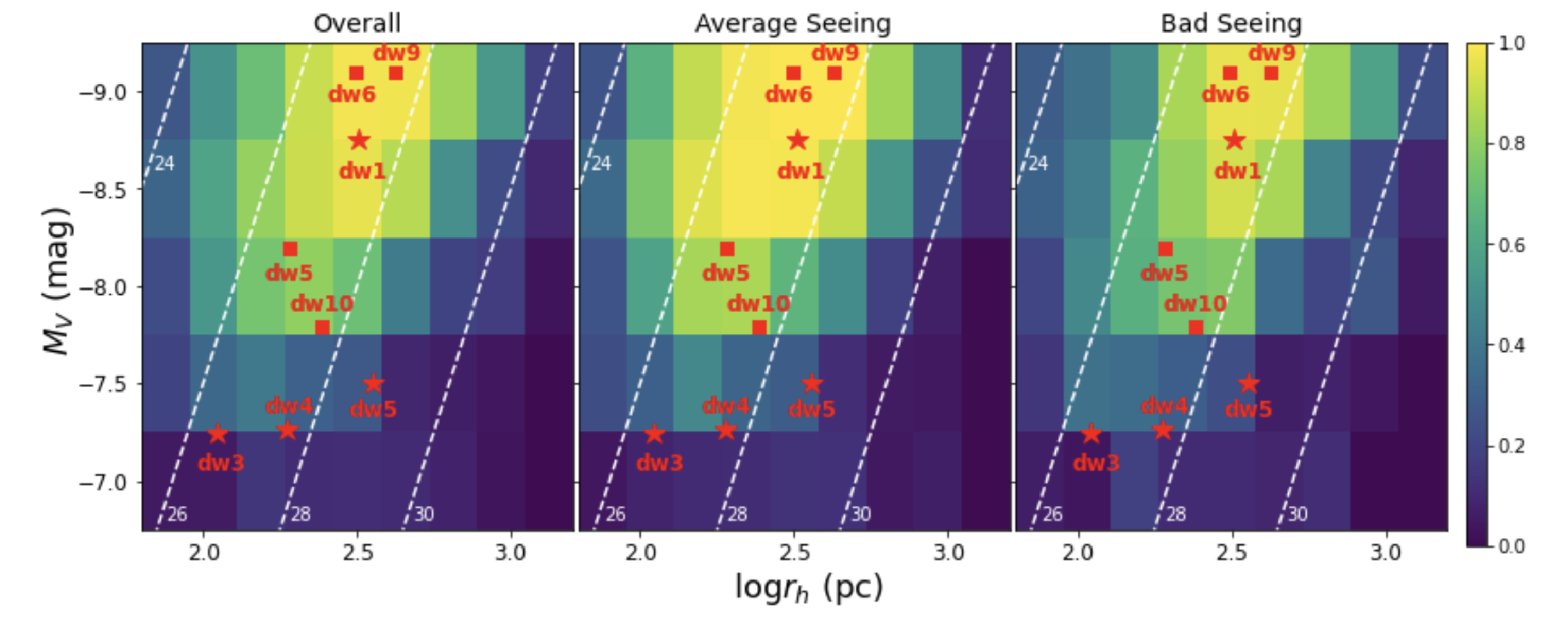}}
\caption{Average completeness of our resolved dwarf searches as quantified with injected artificial galaxies. The left panel shows the results of our tests on 10 representative PISCeS fields, while the middle and right panels focus on the results from the fields with average ($\sim0.8$\arcsec) and bad ($\sim1.0$\arcsec) seeing, respectively. Lines of constant V-band surface brightness are shown at 24, 26, 28, and 30 mag arcsec$^{-2}$. Cen~A dwarfs (red squares) and NGC~253 dwarfs (red stars) are shown as a reference. The color map represents the detection efficiency. 
\label{fig:completeness}}
\end{figure*}

Before comparing our sample to known satellite populations, it is critical to understand and quantify the completeness of the detection algorithm. We perform completeness tests by injecting artificial resolved dwarf galaxies of known magnitude and size, and checking the recovery efficiency. We use the same pipeline to create mock dwarf observations as described in \citet{Mutlupakdil21}. Star positions are drawn randomly from exponential profiles with sizes and luminosities that span those of the detected dwarfs, ellipticities of $<$$0.75$ and position angles ranging from 0 to 360 degrees. We focus on 10 PISCeS fields, 4 with average seeing ($\sim$$0.8$\arcsec) and 6 with bad seeing ($\sim$$1.0$\arcsec) to represent our survey footprint. We inject a total of 5,460 mock galaxies into our coadd-level images, with minimum $\sim$$100$ galaxies per bin in total magnitude and size space. In each iteration, four galaxies are randomly placed in a given field, one for each quadrant. The images with simulated dwarf galaxies are treated the same as the unaltered images, and they are processed through our photometry and detection pipeline.

Figure~\ref{fig:completeness} shows the size-luminosity space probed by our tests, where the colored blocks present the detection efficiency map of our simulated dwarfs. The left panel shows the overall recovery fraction across our representative 10 fields. The center/right panel shows the average recovery fraction of the cases with average/bad seeing. Cen~A dwarfs (red squares) and NGC~253 dwarfs (red stars) are shown as references. 

Our tests clearly show that our overall dwarf search is complete down to $M_V$$\sim$$-8$ and $\mu_{V}$$\sim$$28$~mag arcsec$^{-2}$, with a recovery rate of $>95$\%. There is a clear drop-off in completeness at $\mu_{V}$$\sim$$29$~mag arcsec$^{-2}$, with a $\sim50$\% recovery rate. At $M_V\approx-7.5$, the detectability is about $30$$-$$40$\% for systems with $\mu_{V}$ brighter than $29$~mag arcsec$^{-2}$. This is the luminosity-size space where the known PISCeS UFDs (Scl-MM-dw3, Scl-MM-dw4, and Scl-MM-dw5) are located, and this detectability rate is also consistent with Scl-MM-dw3 being found but the other two being undetected in matched-filter stellar density maps. However, it is worth emphasizing that even for these faint dwarfs, the census of satellites in the PISCeS data is $30$$-$$40$\% complete. 

\section{Beyond the PISCeS footprint}\label{sec:beyond}
We have obtained deep {\it HST} follow-up observations of four dwarf candidates that were discovered beyond the PISCeS footprint \citep{Martinez-Delgado21, Carlsten2022}: Do~III, Do~IV, and dw0036m2828 are confirmed to be satellites of NGC~253, while SculptorSR (ScuSR) is found to be a background galaxy (see Appendix). In this section, we present {\it HST} observations of Do~III, Do~IV, and dw0036m2828. The first two were found by visual inspection of the DESI Legacy Imaging Surveys \citep{Martinez-Delgado21}, while dw0036m2828 was detected from integrated light in the Dark Energy Camera Legacy Survey (DECaLS) using a semiautomated algorithm  \citep{Carlsten2022}. The discovery data are too shallow to constrain their distance and thus confirm their status as NGC~253 satellites. Here, we use {\it HST} follow-up observations that reach $\sim$$3$ magnitudes below the TRGB for each object to derive their distances, structural parameters, and luminosities. 

\subsection{HST Observations and Photometry}\label{sec:hst}
We obtained {\it HST} follow-up observations of these dwarfs (GO-17164, PI: Mutlu-Pakdil) with the Wide Field Channel (WFC) of the Advanced Camera for Surveys (ACS). Each target was observed for a total of one orbit in the F606W and F814W filters: $1006$~s in F606W, $1027$~s in F814W for Do~III, dw0036m2828, and ScuSR; $999$~s in F606W, $1026$~s in F814W for Do~IV.

We performed point-spread function photometry on the pipeline-produced FLC images with the latest version (2.0) of DOLPHOT \citep{Dolphin2000}. We followed the recommended preprocessing steps and used the suggested input parameters from the DOLPHOT User Guide\footnote{\url{http://americano.dolphinsim.com/dolphot/dolphotACS.pdf}}. The initial photometry is culled with the following criteria: the sum of the crowding parameters in the two bands is $<$$1$, the squared sum of the sharpness parameters in the two bands is $<$$0.075$, and the signal-to-noise ratio is $>$$4$ and object-type is $\leq$$2$ in each band. We corrected for MW extinction on a star-by-star basis using the \citet{Schlegel98} reddening maps with the coefficients from \citet{Schlafly11}. The extinction-corrected photometry is used throughout this work.

\begin{figure*}
\centering
\includegraphics[width=0.31\linewidth]{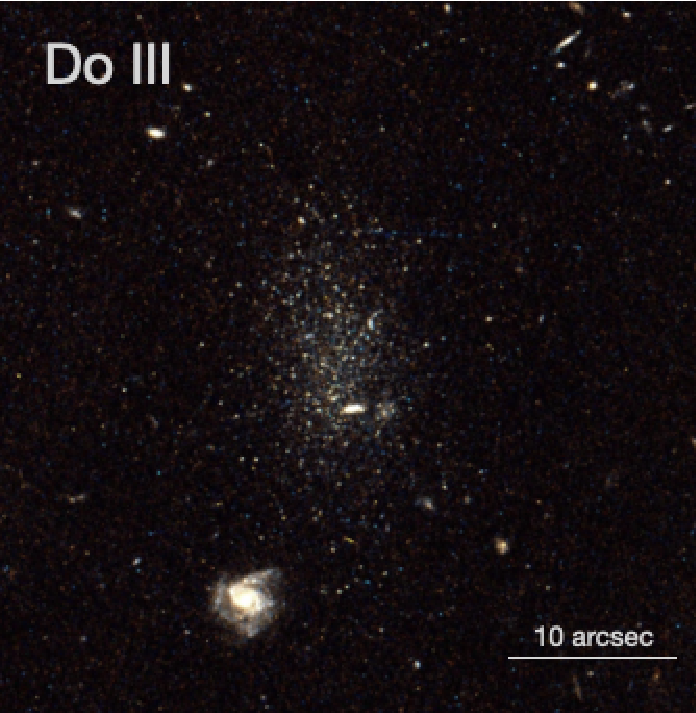}
\includegraphics[width=0.31\linewidth]{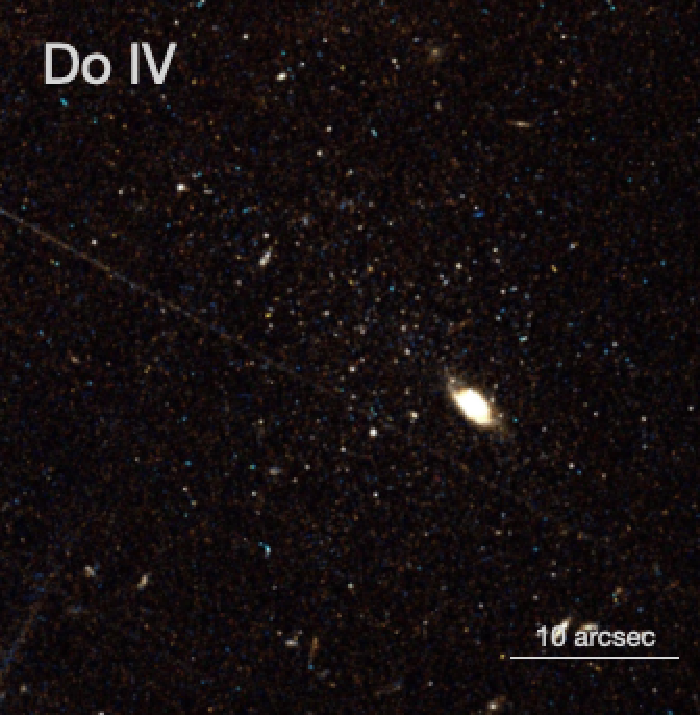}
\includegraphics[width=0.31\linewidth]{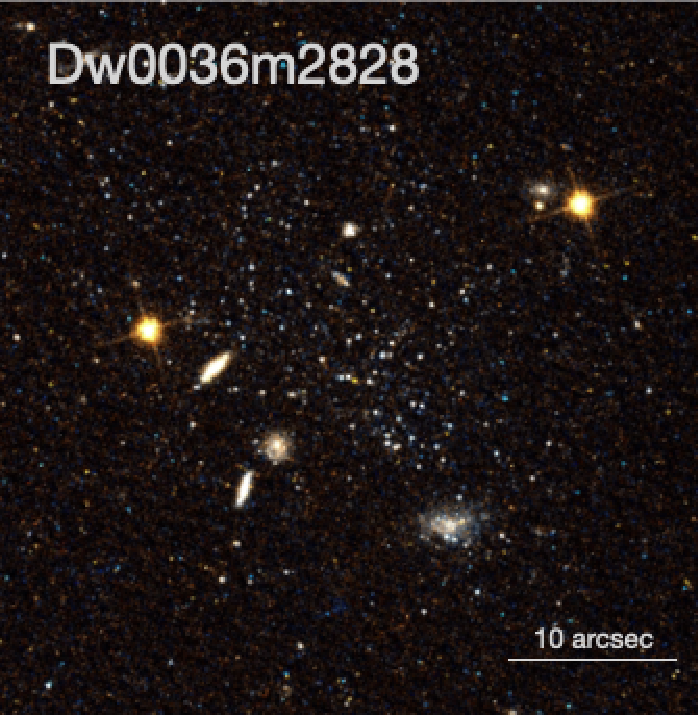}
{\includegraphics[width=\linewidth]{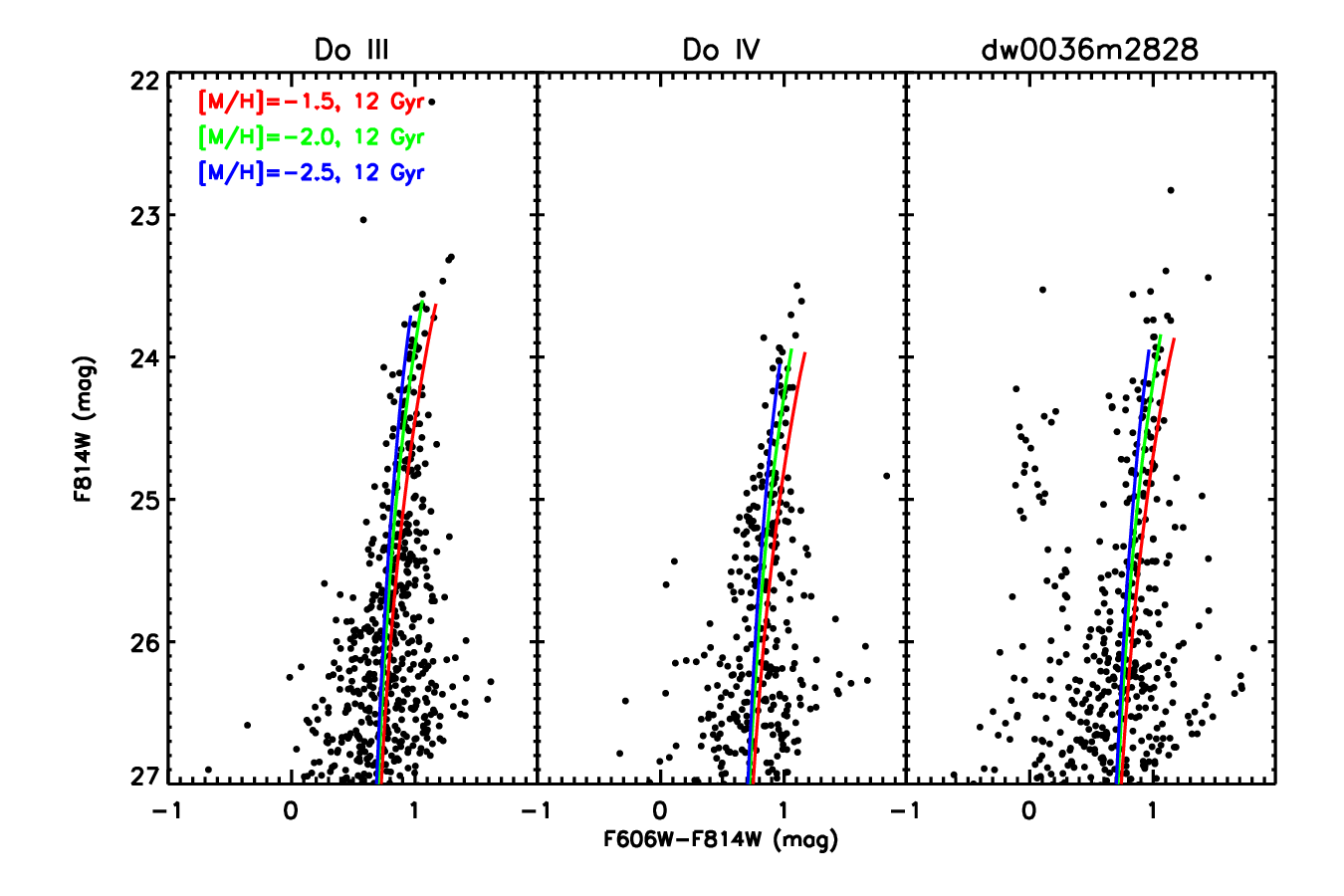}}
\caption{Top: RGB false color {\it HST}/ACS images of Do~III, Do~IV, and dw0036m2828. Bottom: {\it HST} CMDs showing the stars within $2 \times r_h$ of each dwarf galaxy. The blue, green, and red lines indicate the Dartmouth isochrones for 12 Gyr and [M/H] = -2.5 dex, -2.0 dex, and -1.5 dex, respectively. We shift each isochrone by the best-fit distance modulus that we derive using the TRGB in Section~\ref{sec:dist}. \label{fig:cmd}}
\end{figure*}

We performed artificial star tests to quantify the photometric uncertainties and completeness in our observations. A total of $\sim$100,000 artificial stars, implanted one star at a time using the artificial star utilities in DOLPHOT, were distributed uniformly both in color-magnitude space (i.e., $20\leq$F606W$\leq30$ and $-0.5\leq$F606W$-$F814W$\leq1.5$) and spatially across the field of view. Photometry and quality cuts were performed in an identical manner to those performed on the original photometry. Our {\it HST} data are 50\% (90\%) complete at F606W$\sim$27.2~(26.2)~mag and F814W$\sim$26.4~(25.6)~mag.

\subsection{Properties of Do~III, Do~IV, and dw0036m2828}

\subsubsection{Color-Magnitude Diagram}

Figure~\ref{fig:cmd} shows the RGB false color {\it HST}/ACS images of Do~III, Do~IV, and dw0036m2828 (top panel), and their CMDs (bottom panel), which include stars within two half-light radii (see Table~\ref{tab:dwarfs} and Section~\ref{sec:structure}). Overplotted on the CMDs as blue, green, and red lines are the Dartmouth isochrones \citep{Dotter2008} for 12~Gyr and [M/H]$=-2.5$~dex, $-2.0$~dex, and $-1.5$~dex, respectively. Each dwarf is clearly resolved into its constituent RGB stars in the {\it HST} data, and shows old, metal-poor stellar populations at the distance of NGC~253 (see Section~\ref{sec:dist}). The CMDs of Do~III and Do~IV closely resemble that of Scl-MM-dw1 ($M_V=-8.75\pm0.11$, \citealt{MutluPakdil22}), showing a predominantly old stellar population with only a handful of younger asymptotic giant branch (AGB) stars. Due to the relatively low stellar mass of these systems and the inherent unpredictability of the AGB phase, it is difficult to constrain the amount of possible intermediate-age star formation in Do~III and Do~IV. 

\begin{figure}[!ht]
\centering
{\includegraphics[width=\linewidth]{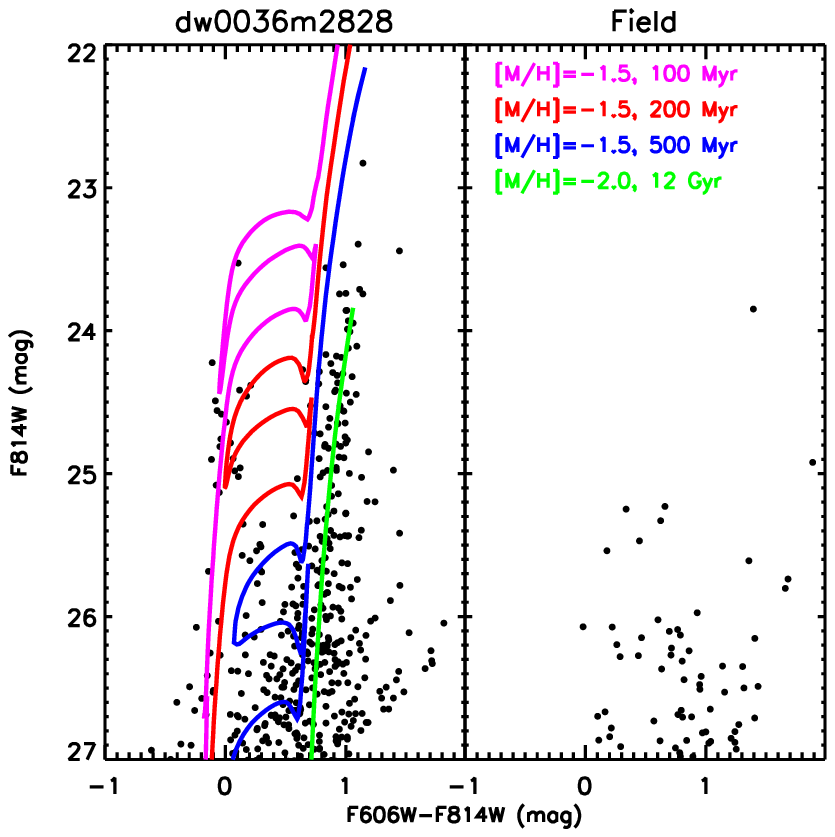}}
\caption{Left: {\it HST} CMDs of dw0036m2828 with PARSEC isochrones for different stellar population ages overlaid. Right: CMD of a representative field region of equal area far away from dw0036m2828. The field region lacks the blue stars consistent with the magenta and red isochrones seen in the CMD on the left. \label{fig:cmd_iso}}
\end{figure}

Unlike other NGC~253 PISCeS dwarfs, dw0036m2828 contains a number of blue stars with F606W$-$F814W$< 0.5$. In Figure~\ref{fig:cmd_iso}, the left panel shows the CMD of dw0036m2828 with PARSEC isochrones \citep{Bressan2012} for different stellar population ages overlaid. These blue stars are consistent with young stellar populations ranging in age from 100 to 500~Myr. The right panel shows a CMD of a field region of equal size located far from the dwarf. This CMD serves as a typical field area in the vicinity of the dwarf, aiding in the assessment of potential field contamination. Note that the field CMD does not contain any stars consistent with the magenta and red isochrones. This provides clear evidence for young stellar populations in dw0036m2828, and we describe our search for an \hi~ reservoir and star-forming gas in this dwarf in Section~\ref{sec:hi}. Deeper {\it HST} or {\it JWST} observations are required to derive detailed star formation histories of dw0036m2828. 

\subsubsection{Distance \label{sec:dist}}
We measure distances to our targets using the TRGB method \citep[e.g.,][]{Lee1993,Salaris2002,Rizzi2007}, as described in \citet{Crnojevic19} and \citet{MutluPakdil22}. We first apply a color correction to our photometry, following \cite{jang17} (their formula 5 and Table 6); we then compute the observed luminosity function for RGB stars, applying a color cut of F606W$-$F814W$>0.6$ to avoid any contamination from possible young populations (see Figure~\ref{fig:cmd_iso}). The luminosity function is fit with a model which is convolved by the appropriate photometric uncertainty and completeness function as derived from our artificial star tests. Our final uncertainties combine the fitting uncertainties (which include the artificial star test results), the uncertainties from the TRGB zeropoint calibration and the applied color correction, and an assumed 10\% uncertainty on the adopted extinction value, added in quadrature. The TRGB values, the distance moduli, and the distances for our targets are reported in Table~\ref{tab:dwarfs}. The good agreement of TRGB distances with the distance of NGC~253 (e.g., \citealt{Radburn-Smith2011}, who found $m-M =27.70\pm0.07$) firmly establishes their association with NGC~253.

\begin{table*}
\centering
\normalsize
\caption{{\it HST}-derived properties of Do~III, Do~IV, and dw0036m2828 \label{tab:dwarfs}}
\begin{tabular}{lccc}
\tablewidth{0pt}
\hline
\hline
Parameter & Do~III & Do~IV & dw0036m2828 \\
\hline
R.A. (deg) & 17.35240$\pm$0.5\arcsec & 11.76255$\pm$1.1\arcsec & 9.12783$\pm$1.6\arcsec \\
Dec. (deg) & $-$27.34707$\pm$0.9\arcsec & $-$21.68071$\pm$1.2\arcsec &  $-$28.46933$\pm$ 1.2\arcsec \\
F814W$_{\rm TRGB}$ (mag) & $23.63\pm0.09$ & $23.96\pm0.32$ & $23.86\pm0.19$ \\
$m-M$ (mag)           &  $27.64\pm0.11$     & $27.98\pm0.33$   & $27.88\pm0.20$\\  
$D$ (Mpc)             &  $3.37^{+0.16}_{-0.17}$    & $3.94^{+0.55}_{-0.64}$  & $3.76^{+0.32}_{-0.35}$\\ 
$M_{V}$ (mag)         & $-8.91\pm0.14$   & $-8.61\pm0.34$ & $-8.75\pm0.35$\\ 
$\log(M_{HI}/M_{\odot})$ & $\lesssim 6.4$    & $\lesssim 6.4$   & $\lesssim 5.2$ \\
$r_{h}$ (arcsec)      &  12.6$\pm$0.8  & 16.2$\pm$2.1 & 18.38$\pm$2.4\\     
$r_{h}$ (pc)   	      &  206$\pm$14 & 309$\pm$40  & 335$\pm$45 \\  
$\epsilon$           &  $0.47\pm0.05$   & $0.41\pm0.09$ & $0.38\pm0.09$ \\ 
Position Angle (deg)  &  173$\pm$4 & 195$\pm$7  & 56$\pm$9\\
\hline
\end{tabular}
  \begin{tablenotes}
      \small
      \item  R.A.: the Right Ascension (J2000.0). DEC: the Declination (J2000.0). F814W$_{\rm TRGB}$: TRGB magnitude in F814W. $m-M$: the distance modulus. $D$: the distance of the galaxy in Mpc. $M_{V}$: the absolute V-band magnitude. $log(M_{HI}/M_{\odot})$: $3\sigma$ upper limits on the \hi~mass of each object. $r_{h}$: the elliptical half-light radius along the semi-major axis. $\epsilon$: ellipticity which is defined as $\epsilon=1-b/a$, where $b$ is the semiminor axis and $a$ is the semimajor axis.
    \end{tablenotes}
\end{table*}

\subsubsection{Structural Properties \label{sec:structure}}
We derive structural parameters (including half-light radius $r_h$, ellipticity, and position angle) using the maximum-likelihood (ML) method of \citet{Martin08}, as described in \citet{MutluPakdil22}. In our analysis, we select stars consistent with an old, metal-poor isochrone in color-magnitude space after taking into account photometric uncertainties, within our 90\% completeness limit. We inflate the uncertainty to 0.1~mag when the photometric errors are $<0.1$~mag for the purpose of selecting stars to go into our ML analysis. The stellar profiles of the dwarfs are well described by a single exponential model \citep[e.g.,][]{Martin08,Munoz18}. We fit a standard exponential profile plus constant background to the data, and summarize the resulting structural parameters in Table~\ref{tab:dwarfs}. Uncertainties are determined by bootstrap resampling the data 1500 times and recalculating the structural parameters for each resample. 

\citet{Martinez-Delgado21} reported the structural parameters of Do~III and Do~IV, derived with integrated light from DESI Legacy imaging: $r_h=8.46\pm0.17$~arcsec with $\epsilon=0.41\pm0.01$ for Do~III and $r_h=10.21\pm0.65$~arcsec with $\epsilon=0.40\pm0.03$ for Do~IV. In comparison, our ML analysis suggests larger sizes with similar elongated shapes: $r_h=12.6\pm0.8$~arcsec with $\epsilon=0.47\pm0.05$ and $r_h=16.2\pm2.1$~arcsec with $\epsilon=0.41\pm0.09$ for Do~III and Do~IV, respectively. We note that \citet{Martinez-Delgado21} performed a Sersic profile fitting, and obtained Sersic index $n=0.56$ for Do~III and $0.85$ for Do~IV, respectively. As we employ an exponential profile ($n=1$), the size difference might be due to the difference in Sersic index.    

\subsubsection{Luminosity \label{sec:lum}}
We derive absolute magnitudes by using the same procedure as in \citet{MutluPakdil22}, as was first described in \citet{Martin08}. First, we create a well-populated CMD (of $\sim$20,000 stars) in {\it HST} filters, including our completeness and photometric uncertainties, by using the Dartmouth isochrone with age 12~Gyr and [M/H]$=-2.0$~dex and its associated luminosity function assuming a Kroupa IMF \citep{Kroupa2001}. We randomly select the same number of stars from this artificial CMD as was found from our profile fits. We obtain the total luminosity by summing the flux of these stars, and extrapolating the flux of the faint, unresolved component of the galaxy from the adopted luminosity function. We perform 1000 realizations in this way, and take the mean as our absolute magnitude and its standard deviation as the uncertainty. To address the uncertainty in the number of stars (assuming Poisson statistics), we perform this procedure 100 times by adjusting the number of stars within its uncertainty range, and use the offset from the best-fit value as the associated uncertainty. These error terms and the distance modulus uncertainty are then added in quadrature to produce our final uncertainty on the absolute magnitude. 

We find $M_V$$=$$-8.91\pm0.14$~mag for Do~III and $M_V$$=$$-8.61\pm0.34$~mag for Do~IV. Our Do~III result is consistent with the value reported in the initial discovery analysis within the uncertainties, whereas our Do~IV value is significantly brighter \citep[$M_V$$=$$-9.13\pm0.09$~mag for Do~III and $M_V$$=$$-7.89\pm0.15$~mag for Do~IV;][]{Martinez-Delgado21}. 

For dw0036m2828, we also account for the luminosity contribution from its young stellar populations. We focus on stars with F814W $<$$26$~mag and F606W-F814W $<$$0.4$ within 2$\times$$r_{h}$, and calculate the total flux emitted by these stars. Given the overall agreement between these blue stars and the red-labeled isochrone shown in Figure~\ref{fig:cmd_iso} (the one with an age of 200~Myr), we adopt the corresponding PARSEC luminosity function and extrapolate the flux of unaccounted young stars. The sum of the total luminosity of old RGB stars and these young stars yields an absolute magnitude of $M_V=-8.75\pm0.35$~mag for dw0036m2828\footnote{We also repeat this calculation by assuming two young subpopulations: one with an age of 200~Myr (applying to stars with F814W $< 25.2$ and F606W-F814W $< 0.4$) and one with an age of 500~Myr (applying to stars with $25.2<$F814W $<26$ and F606W-F814W $<0.4$). This gives $M_V=-8.79$, showing a close agreement with our reported value.}. 

\subsubsection{\hi~Gas Limits \label{sec:hi}}

The Galactic All Sky Survey \citep[GASS;][]{McClure-Griffiths09,Kalberla10,Kalberla15} contains all three dwarfs and has a bandwidth of approximately $-500 < v/\mathrm{km\,s^{-1}} < 500$. Note that the radial velocities of NGC~253 and NGC~247 are $261~\mathrm{km\,s^{-1}}$ and $153~\mathrm{km\,s^{-1}}$, respectively. 
We downloaded \hi \ line emission cubes from GASS\footnote{\url{https://www.astro.uni-bonn.de/hisurvey/gass/index.php}}, but saw no signs of emission at the location of any of our dwarf targets. The rms noise in these cubes is 50~mK in 0.82~km~s$^{-1}$ channels. If we approximate the Parkes telescope gain as 0.7~K/Jy, then this corresponds to $\sim$70~mJy. Assuming a fiducial velocity width of 20~km~s$^{-1}$ for the dwarfs, this gives a 3$\sigma$ detection limit of $\log(M_\mathrm{HI}/M_{\odot}) \lesssim 6.4$ for all three dwarfs. This limit is consistent with the dwarfs being gas-poor, however, we note that it is not a strong limit. For example, the star-forming, low-mass dwarf Leo~P ($M_V$$=$$-9.4$, \citealt{Giovanelli13,McQuinn2015}) would be undetected in these data if it were at the distance of NGC~253. In addition, a notable caveat to these limits is that if the radial velocities of the dwarfs are very small (e.g. $|v| < 100$~km~s$^{-1}$), then any \hi \ emission could still be blended with the MW.

One of the dwarfs (dw0036m2828) was observed in October 2023 as part of the Green Bank Telescope (GBT) program GBT23A-084 (PI: M.~Jones). With approximately 1~h of on/off integration an rms noise of 1.7~mJy was achieved at 5~km~s$^{-1}$ resolution. There is a $\sim$2.5$\sigma$ peak in the spectrum at a velocity of $\sim$300~km~s$^{-1}$, however, when the spectrum is split into its two component polarizations they strongly disagree at this velocity. Thus, this is almost certainly a spurious signal. Following an equivalent process to that above, we obtain a 3$\sigma$ detection limit of $\log(M_\mathrm{HI}/M_{\odot}) \lesssim 5.2$ for dw0036m2828. In this case, if the radial velocity of dw0036m228 is in the range $-200 < v < 100$~km~s$^{-1}$ then this limit does not apply due to contamination from the MW.
\vspace{10mm}

\subsection{GALEX UV Imaging}

Data from GALEX \citep[GAlaxy Evolution EXplorer;][]{galex2005} were used to measure the star formation rate (SFR) for dw0036m2828, which was observed for 224~s in both FUV and NUV as part of the GALEX All-Sky Imaging Survey. The UV is a good tracer for star formation with the GALEX FUV corresponding to star formation in the past $\sim$10~Myr with a tail of response out to $\sim$100~Myr, while GALEX NUV corresponds to star formation in the past $\sim$100~Myr with a tail of response out to $\sim$250~Myr \citep{Calzetti_2013}. Recent work has shown that SFR derived from GALEX UV emission agrees well with that found in H$\alpha$ for low mass dwarf galaxies \citep{Jones2023b}. 

dw0036m2828 is visible in GALEX. To obtain a UV magnitude we use aperture photometry, with the aperture set to twice the half-light radius found in \S \ref{sec:structure}. We sum all flux in this aperture, after masking background sources, yielding apparent magnitudes of m$_{NUV}$~$=$~20.6$\pm$0.3 and m$_{FUV}$~$=$~24.3$\pm$0.5. We then convert to the absolute magnitude to derive the SFR using the relations from \citet{IglesiasParamo2006}. This yields SFR$_{NUV}$~$=$~3.1$\pm$0.8$\times$10$^{-5}$~M$_{\odot}$ yr$^{-1}$ and SFR$_{FUV}$~$=$~5.1$\pm$2.4$\times$10$^{-7}$~M$_{\odot}$ yr$^{-1}$. This suggests a low level of star formation over the past $\sim$100~Myr, with a higher level over the past $\sim$250~Myr, this matches with what we see in the CMD (see Figure~\ref{fig:cmd_iso}). 

Examination of Do~III in GALEX data showed no evidence of UV emission, this is consistent with the CMD which shows no young stellar population. Do~IV is in a gap in the GALEX footprint, but from the CMD we would expect no UV emission. Similarly, the analysis of other PISCeS NGC~253 dwarfs, specifically Scl-MM-dw1, Scl-MM-dw3, and Scl-MM-dw5, showed no signs of UV emission, as expected by their HST CMDs \citep{MutluPakdil22}. Note that Scl-MM-dw2 and Scl-MM-dw4 are in a gap in the GALEX coverage, and based on their CMDs, no UV emission is expected.

\subsection{Comparison to Known Local Volume Dwarfs}

\begin{figure}[!h]
\centering
\includegraphics[width = \linewidth]{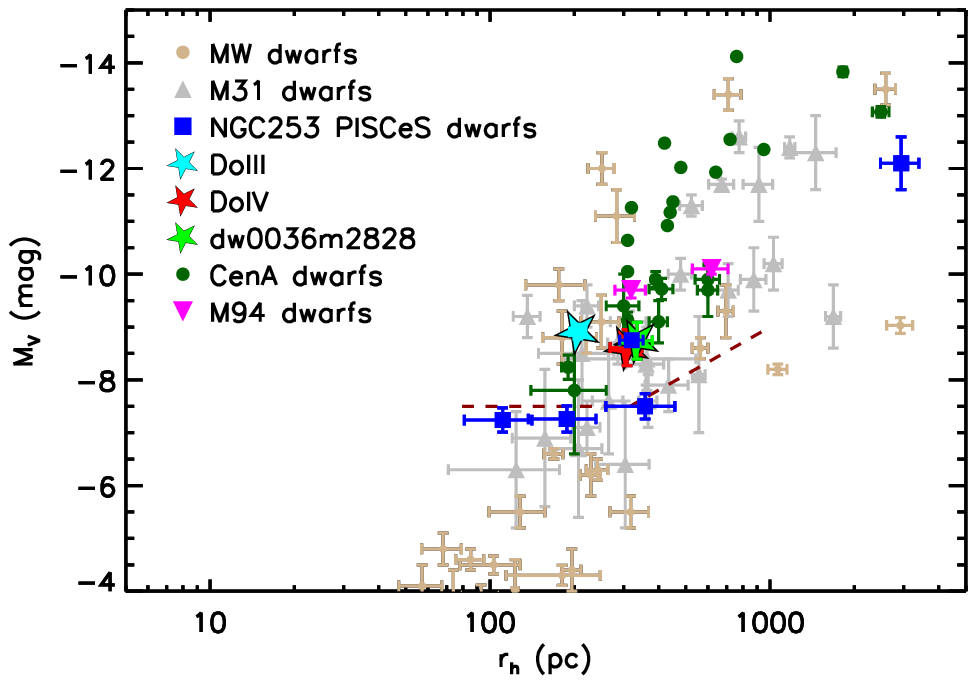}
\caption{Absolute V-band magnitude as a function of half-light radius for Do~III (cyan star), Do~IV (red star), and dw0036m2828 (green star), relative to MW/M31 dwarf galaxies, Cen~A dwarfs \citep[][note that the uncertainties for several Cen~A dwarfs were not provided]{Sharina2008,Crnojevic19}, and M94 dwarfs \citep{Smercina18}. The dashed line represents the $\sim$$50$\% completeness limit in PISCeS. All three have similar properties to those of LG dwarfs and known Local Volume dwarfs. \label{fig:compare}}
\end{figure}

Figure~\ref{fig:compare} shows Do~III, Do~IV, and dw0036m2828 in the size-luminosity plane, relative to the Local Group dwarfs, NGC~253 dwarfs from our PISCeS program, as well as Cen~A and M94 dwarfs. All three are comparable to known Local Volume dwarf galaxies. Our luminosity measurements place them near the faint end of those of the classical dSphs in the MW and M31. The MW satellite most similar to them is Draco ($M_V=-8.8\pm0.3$~mag; $r_h=221\pm26$~pc; $\epsilon=0.31\pm0.02$, \citealt{McConnachie2012}). Among the NGC~253 PISCeS dwarfs, the closest analog is Scl-MM-dw1 ($M_V=-8.75\pm0.11$~mag; $r_h=321\pm31$~pc; $\epsilon=0.20\pm0.07$, \citealt{MutluPakdil22}), with a few luminous asymptotic giant branch (AGB) stars. While both Do~III and Do~IV do not seem to contain populations younger than $\sim6-8$~Gyr (similar to Scl-MM-dw1), dw0036m2828 has young stellar populations ranging in age from 100 to 500 Myr. Given the comparable projected distance to NGC~253, this might suggest that Do~III and Do~IV have already experienced significant environmental processing by a pericentric passage while dw0036m2828 is likely on its first infall into the system.  
%\vspace{2mm}

\section{Plane of Satellites in NGC~253?}\label{sec:plane}

\citet{Martinez-Delgado21} recently suggested the existence of a plane of satellites around NGC~253. This thin satellite structure was based on seven objects (Scl-MM-dw1, Scl-MM-dw2, LVJ0055-2310, DDO~6, NGC~247, ESO~540-032, and KDG2), excluding the four candidates which did not have any distance measurements (Do~III, Do~IV, Scl-MM-dw3\footnote{We note that \citet{Martinez-Delgado21} published the independent discovery of one of our PISCeS dwarfs, and named it as Do~II}, ScuSR). Since then, three more new dwarf galaxies have been discovered (two of them from our PISCeS data, \citealt{MutluPakdil22}: Scl-MM-dw4 and Scl-MM-dw5; one from the ELVES Survey, \citealt{Carlsten2022}: dw0036m2828). We use deep {\it HST} observations to confirm their membership with NGC~253 by deriving accurate TRGB distances: all but ScuSR are indeed a part of the NGC~253 dwarf satellite system (see \citealt{MutluPakdil22}; and Section~\ref{sec:beyond} and Appendix~\ref{sec:appendix} of this paper). We revisit the spatial distribution of satellites in light of this new information. 

\begin{figure}
\centering
\includegraphics[width = \linewidth]{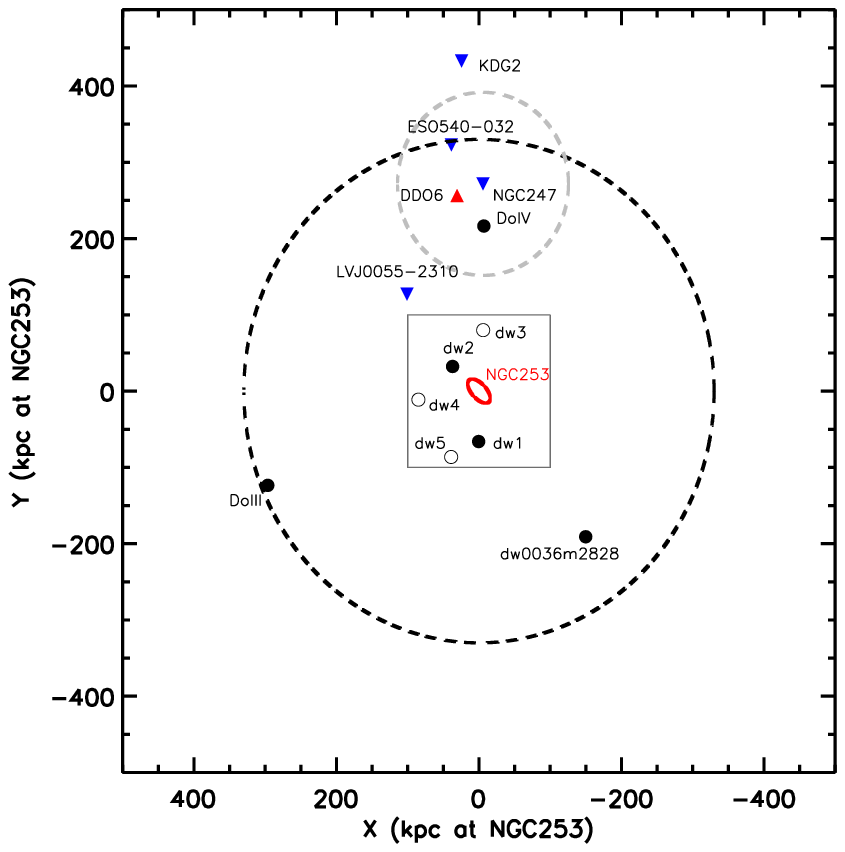}
\caption{The location of confirmed satellites relative to NGC~253. The PISCeS footprint is shown with a square, and known UFDs are indicated with open circles. Dwarfs without velocity measurements are shown with filled circles while ones with known velocities are color-coded for approaching (blue, downward triangles) and receding (red, upward triangles) systems, according to their line-of-sight velocities relative to the NGC~253 velocity. The black dashed circle outlines the approximate virial radius of NGC~253 (330~kpc), while the grey dashed one represents the approximate virial radius of NGC~247 (120~kpc, \citealt{Mutlupakdil21}; $D=3.7$~Mpc, see Table~\ref{tab:members}). The proposed plane is oriented in the north-south direction. However, the asymmetric distribution of satellites in NGC~253 can be explained by the presence of an NGC~247 sub-group.    
\label{fig:plane}}
\end{figure}

Figure~\ref{fig:plane} shows the location of satellites relative to NGC~253. The PISCeS UFDs are shown with open circles to emphasize that such faint systems require deep imaging, such as the PISCeS dataset, and thus they are highly incomplete beyond our footprint, which is depicted with a square. We also marked the approximate virial radius of NGC~247 (NGC~253's most massive satellite), which has a stellar mass similar to the Large Magellanic Cloud (LMC). The $\Lambda$CDM model predicts that even moderate-sized dwarf galaxies should host their own satellites \citep{Munshi19}. Recent observational programs have revealed the spatial clustering of dwarf galaxies in the vicinity of LMC \citep[e.g.,][]{Bechtol2015,Drlica-Wagner2015,Koposov2015}. There is a significant ongoing effort to map out the halos of several low-mass galaxies and search for their satellite populations, e.g., PandAS around M33 (with CFHT/MegaCam, two possible satellites have been reported; \citealt{Martin2009,Martin2013,Martinez-Delgado2022}), the MADCASH Survey (a DECam$+$HSC deep imaging campaign around a dozen isolated nearby low-mass galaxies, where three satellites have been reported: one around NGC~2403, one around NGC~4214, and one around NGC~3109; \citealt{Sand15b,Carlin16,carlin21}), and DELVE-DEEP (with DECam, one satellite has been reported around NGC~55; \citealt{McNanna2023}). Therefore, it is not surprising to observe a similar clustering of dwarf galaxies in the vicinity of NGC~247. Furthermore, the confirmed memberships of Do~III and dw0036m2828, which are significantly offset from the suggested thin satellite structure, argue against the existence of a satellite plane. Given that these two satellites fall below the completeness limit of $M_V\approx-9$, there could be other similar systems awaiting discovery. As PISCeS has shown, there are likely more ultra-faint satellites within the virial volume of NGC~253. Considering that the suggested flattened distribution by \citet{Martinez-Delgado21} did not show the same degree of tension as those around the MW, M31, and Cen~A, our new observations render the NGC~253 system more typical compared to the expectations derived from cosmological simulations. 
   
In Figure~\ref{fig:plane}, five dwarfs with known velocities are color-coded for approaching (blue, downward triangles) and receding (red, upward triangles) systems relative to NGC~253 (see Table~\ref{tab:members}). The number of galaxies with known velocities is currently too small to suggest any coherent rotation. As also pointed out by \citet{Martinez-Delgado21}, the velocity of NGC~247 is relatively different than that of the other four galaxies. These dwarfs could be spatially and kinematically related, but are not likely to be in dynamical equilibrium, even if they are gravitationally bound (as in dwarf associations in \citealt{Tully2006}). To investigate this further, it is necessary to conduct spectroscopic follow-up studies of dwarf galaxies lacking measured velocities. 
 
In short, we find no strong evidence for the presence of a plane of satellites around NGC~253, and suggest that the asymmetric distribution of satellites in NGC~253 can be explained by the presence of an NGC~247 sub-group, a natural expectation.
\vspace{10mm}

\section{NGC~253 Satellite Luminosity Function}\label{sec:LF}

\begin{figure}[!b]
\centering
\includegraphics[width = \linewidth]{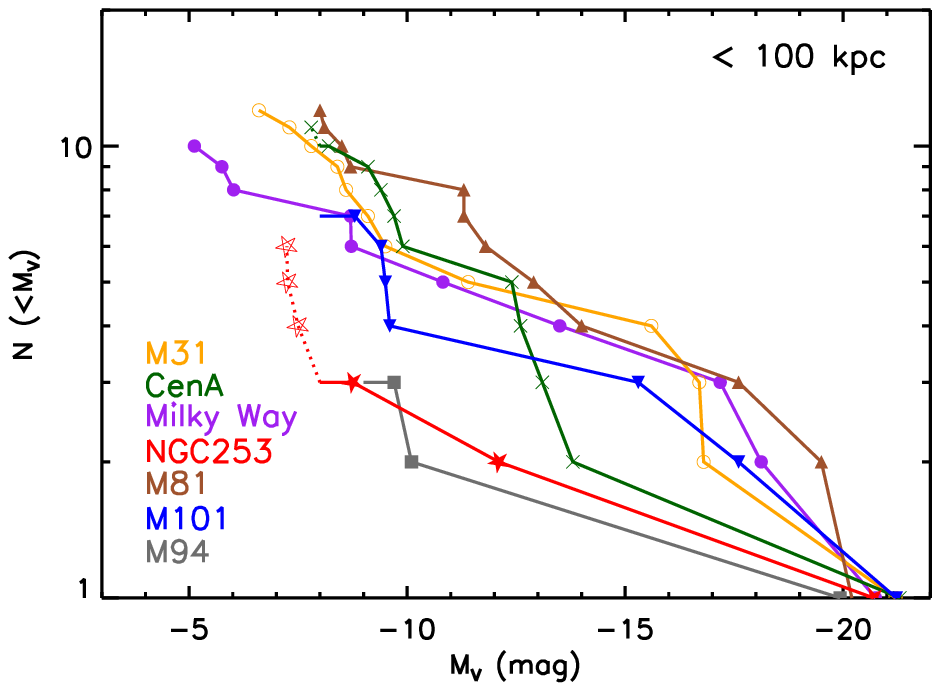}
\includegraphics[width = \linewidth]{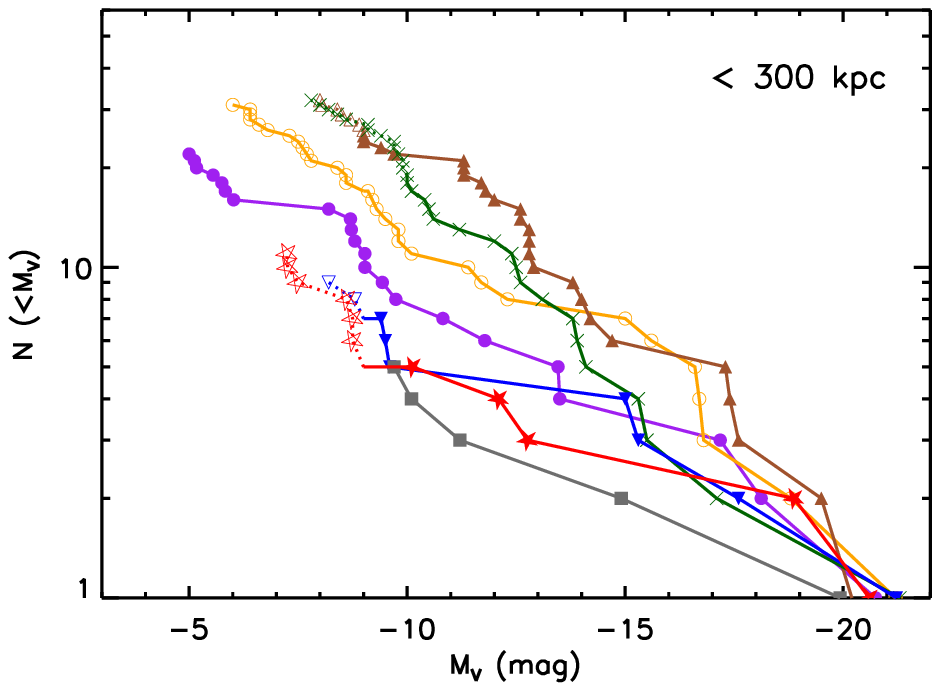}
\caption{The cumulative satellite LFs for NGC~253 (red stars), the MW (purple filled circles), M31 (yellow open circles), Cen~A (green crosses), M81 (brown triangles), M101 (blue upside down triangles), and M94 (gray squares). The top panel shows satellites within a projected radius (or 3D radius for the MW) of 100~kpc; the bottom panel includes objects within 300~kpc of each host. No attempt was made to correct any LF for incompleteness. We denote the region where the LFs of NGC~253, M101, and M81 become incomplete with hollow symbols and dashed lines. Galaxies are listed in descending order of stellar mass. \label{fig:slf}}
\end{figure}

\begin{table*} 
\centering
\normalsize
\caption{Galaxies in the vicinity of NGC~253; those located beyond the PISCeS footprint are below the horizontal line \label{tab:members}}
\begin{tabular}{lcccccccc}
\tablewidth{0pt}
\hline
\hline
Galaxy & R.A.  & Dec  & $M_V$ & $V_h$ & $D_{\text{TRGB}}$  & $D_{\text{proj}}$ & $D_{\text{3D}}$  & References \\
 & (deg) &  (deg) &  (mag) & (km s$^{-1}$) & (Mpc) & (kpc)  &  \\
(1)    &    (2)     & (3)         & (4)         &  (5)                    & (6)       & (7)    & (8) & (9)   \\
\hline
NGC~253    & 11.88800 & -25.28822 & $-20.60$ & $261\pm5$ &  $3.50\pm0.1$ & \nodata & \nodata & 3,4 \\
Scl-MM-dw1 & 11.89643 & -26.38971 & $-8.75\pm 0.11$ & \nodata & $3.53\pm0.55$ & 66 & 74 & 1 \\
Scl-MM-dw2 & 12.57108 & -24.74961 & $-12.10\pm 0.50$ & \nodata & $3.53\pm0.11$ & 50 & 58 & 1,2 \\
Scl-MM-dw3 & 11.77950 & -23.95573 & $-7.24^{+0.26}_{-0.21}$ & \nodata & $3.48^{+0.14}_{-0.28}$ & 81 & 84 & 1 \\
Scl-MM-dw4 & 13.45476 & -25.47442 & $-7.26^{+0.27}_{-0.23}$ & \nodata & $4.10^{+0.16}_{-0.32}$ & 86 & 606 & 1 \\
Scl-MM-dw5 & 12.60776 & -26.72726 & $-7.50^{+0.28}_{-0.20}$ & \nodata &$3.90^{+0.18}_{-0.27}$ & 96 & 411 & 1 \\
\hline
LVJ0055-2310 & 13.75456 & -23.16880 & $-10.12$ & $250\pm5$ & $3.62\pm0.18$ & 166 & 204 & 5,6 \\
Do IV & 11.76255 & -21.68071 & $-8.61\pm0.34$ & \nodata & $3.94^{+0.55}_{-0.64}$ & 220 & 492 & 0 \\
dw0036m2828 & 9.12783 & -28.46933 & $-8.75\pm0.35$ & \nodata & $3.76^{+0.32}_{-0.35}$ & 247 & 359 & 0 \\
DDO~6 & 12.45500 & -21.01500 & $-12.76$ & $295\pm5$ & $3.44\pm0.15$ & 263 & 270 & 5,7 \\
NGC 247 & 11.78562 & -20.76039 & $-18.87$ & $153\pm5$ & $3.72\pm0.03$ & 277 & 353 & 5,7 \\
Do~III & 17.35240 & -27.34707 & $-8.91\pm0.14$ & \nodata & $3.37^{+0.16}_{-0.17}$ & 346 & 478 & 0 \\
ESO 540-032 & 12.60134 & -19.90672  & $-11.76$ & $228\pm1$ & $3.63\pm0.05$ & 327 & 352 &  7,8 \\
KDG2 & 12.33734 & -18.07542 & $-11.80$ & $224\pm3$ & $3.56\pm0.07$ & 441 & 445 & 7,8 \\
\hline
\end{tabular}
  \begin{tablenotes}
      \small
      \item  (1) Galaxy name; (2) the Right Ascension (J2000.0); (3) the Declination (J2000.0); (4) the absolute V-band magnitude; (5) heliocentric velocity in km s$^{-1}$ with its error; (6) TRGB–distance in Mpc with a corresponding error; (7) projected distance in kpc at NGC~253 distance; (7) spatial distance to NGC~253 in kpc; (9) references: (0) this work, (1) \citet{MutluPakdil22}, (2) \citet{Toloba16}, (3) \citet{Radburn-Smith2011}, (4) \citet{Bailin2011}, (5) \citet{Westmeier2017}, (6) \citet{Karachentsev2021}, (7) \citet{Jacobs2009}, (8) \citet{Bouchard2005}. 
    \end{tablenotes}
\end{table*}

While the Local Group will remain an important testing ground for understanding the astrophysics and cosmological implications of the very faintest dwarf galaxy satellites ($M_V \gtrsim -7$, e.g., \citealt{Munshi19,Nadler2021}, among others), the faint satellite LFs of nearby galaxy systems are necessary to provide context to Local Group studies and explore how the LF changes with primary halo mass, environment, and morphology. This motivated a significant observational effort to survey the satellite populations of nearby MW-mass systems through wide-field integrated light searches, targeted resolved star studies, and spectroscopic surveys \citep[e.g.,][]{Danieli17,Bennet19,Geha17,Mao2021,Carlsten2022}. Beyond the Local Group, dedicated deep wide-field surveys exist for Cen~A \citep[][with Magellan/Megacam]{Crnojevic16,Crnojevic19}, M94 \citep[][with Subaru/HSC]{Smercina18}, and M81 (including the M82 region; \citealt{Chiboucas2009,Chiboucas13}, with CFHT/MegaCam; \citealt{Okamoto2019,Bell2022}, with Subaru/HSC). Among these deep surveys, the faintest dwarfs discovered so far (and later confirmed with deeper imaging from {\it HST}) are  CenA-MM17-Dw10 ($M_V = -7.8$, $r_h =$ 250~pc; \citealt{Crnojevic19}) in the Cen~A group, and $d0944+69$ ($M_V = -8.1$, $r_h =$ 130~pc; \citealt{Chiboucas13}) in the M81 group. Our NGC~253 PISCeS program brings the total number of surveyed systems to four, uncovering the first examples of (confirmed) UFD satellites of an MW-mass galaxy beyond the Local Group \citep{MutluPakdil22}.

We compile the LF for the Local Group and for nearby groups of galaxies with satellites confirmed via distance measurements. For the MW, we adopt the \citet{Drlica-Wagner2021} compilation, and we only consider objects with $M_V<-5$ (except for Sagittarius~II, which has $M_V=-5.2$ but was suggested to be a globular cluster, see \citealt{MutluPakdil2018,Longeard2021}). For M31, we use the catalog presented in \citet{Savino2022}, and we also include IC~10 \citep{McConnachie2018} and Peg~V \citep{Collins2022}. The PandAS survey covers the inner projected $150$~kpc volume of M31, and is shown to be sensitive to ultra-faint satellites with $M_V \lesssim -6$ \citep{Amandine2022}. With Pan-STARRS, we assume the census of M31 satellites to be complete out to $\sim300$~kpc down to $M_V \sim -9$ \citep[e.g.,][]{Martin2013a,Martin2013b}. For Cen~A, we use the results from \citet{Crnojevic19} and \citet{Muller2019}. \citet{Crnojevic19} estimate the completeness limit to be at $M_V \sim -8$ over the Cen~A PISCeS footprint (which covers the inner projected $150$~kpc) while \citet{Muller2019} estimate that they are complete down to $M_V \sim -10$ over the inner projected $200$~kpc \citep{Muller2017}. For M81, we utilize the Updated Nearby Galaxy Catalog \citep[UNGC,][]{Karachentsev2013}, and complement it with Table~3 of \citet[][we exclude possible tidal dwarfs]{Chiboucas13}. The approximate completeness limit for M81\footnote{Recently, \citet{Bell2022} reported new UFD candidates in the M81 group, but no completeness limit for their search was reported, and those candidates have not yet been confirmed with {\it HST}.} is at $M_V\sim -8$ throughout the inner projected $250$~kpc volume \citep{Chiboucas13}. We convert the $r$ magnitudes reported in \citet{Chiboucas13} and the $B$ magnitudes in UNGC into $V$ magnitudes, assuming $M_V = M_r + 0.4$ and $M_V = M_B - 0.31$ \citep{Crnojevic19}, respectively. For M94, we include the discoveries from \citet{Smercina18} as well as two distant group members (KK~160 and IC~3687). \citet{Smercina18} estimate the approximate completeness limit for M94 to be at $M_V \sim -9$ throughout the inner projected $150$~kpc volume.  For M101, we adopt the \citet{Bennet20} LF, which is complete down to $M_V \sim -8$ out to $\sim250$~kpc. Finally, we have compiled an updated table of the NGC~253 satellites with projected distances $\lesssim 500$~kpc, including their coordinates, luminosities, velocities, TRGB distances, projected distances, and spatial distances (Table~\ref{tab:members}). We note that our census is complete down to $M_V$$\sim$$-8$ out to 100~kpc from PISCeS, and $M_V$$\sim$$-9$ out to 300~kpc based on ELVES \citep{Carlsten2022}.  

The top panel of Figure~\ref{fig:slf} shows the inner satellite systems, $r_{proj} < 100$~kpc, while the bottom panel shows abundances out to $300$~kpc of each host. Except for the MW, where 3D distances are used, we adopt projected distances. As most deep wide-field surveys (beyond the Local Group) are limited to $\sim150$~kpc (or $250$~kpc at most), the LFs in the bottom panel ($r_{proj} < 300$~kpc) should be considered as a lower limit due to incomplete spatial coverage. None of the reported LFs have been corrected for incompleteness effects, but we denote the region where the LFs of NGC~253, M101, and M81 become incomplete with hollow symbols and dashed lines. Overall, the NGC~253 LF is consistent with the Local Volume sample. It is intriguing that its slope is more similar to those of the relatively isolated M94 and M101 galaxies, suggesting a possible correlation with the surrounding environment.

It is worth noting that, while these area-limited LF measurements may be the only way to achieve a fair comparison, they come with inherent uncertainties. As satellites are not stationary and move around the host in an orbit, adopting a fixed projected distance might result in some of the satellites being somewhat randomly excluded or included in the comparisons. While the standard fiducial virial radius adopted for MW-mass hosts in the literature is $\approx$300~kpc, using a slightly different radius, such as 350~kpc, would result in the inclusion of systems like Do~III and ESO~540-032 in the LF, slightly increasing the slope. Although this would not change any of our conclusions here, it highlights some of the hidden uncertainties that are not often considered when studying the halo-to-halo scatter and reaching a broader view of satellite systems. 

Additionally, we are viewing these systems along a cone and therefore sampling out to a much larger radius along the line of sight. This means that on the sky we may be sampling to the virial radius (or some fraction) but along the line of sight, we also include the infall region and beyond. If everything was scale-free, then perhaps we would always have the same larger factor of `satellites' (i.e. we would always have twice as many satellites as if we were able to cut in virial radius along the line of sight as well). However, this correction is likely halo mass dependent, and will certainly also be affected by environment. The contamination from objects in the infall region can be significant. \citet{Goto2023} investigated how this contamination affected the projected radial density profile, but it also has implications for determining star-forming fractions, gas fractions, etc. It may be, for example, a contributing reason the satellite numbers scale with local density (see next section). The MW and M31 numbers are not susceptible to this problem but more distant systems are.

\section{Discussion}\label{sec:discussion}
In this section, we briefly explore the characteristics and trends among the satellite systems, as a function of the most dominant mergers, host stellar mass, local density environment, and morphology. 

\subsection{Total Satellite Counts}

\begin{figure}[!b]
\centering
\includegraphics[width = \linewidth]{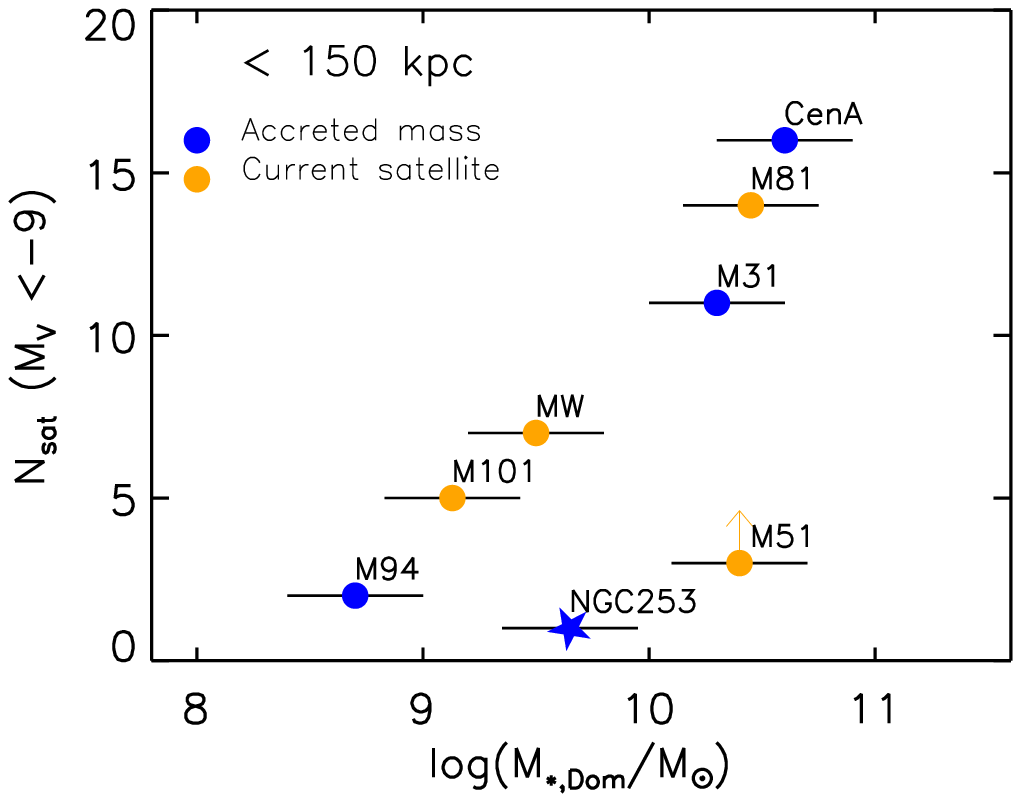}
\caption{Total number of satellites with $M_V < -9$, within 150 kpc projected radius, around each of seven nearby MW-mass galaxies, as a function of the mass of the most dominant merger they have experienced \citep{Smercina2022}. Galaxies are color-coded according to whether $M_{\star,Dom}$ reflects the accreted material from a past merger (blue), or the mass of an existing satellite (orange). NGC~253 is a clear outlier in this linear relationship. \label{fig:smercina}}
\end{figure}

\begin{figure*}[!t]
\centering
\includegraphics[width = 0.475\linewidth]{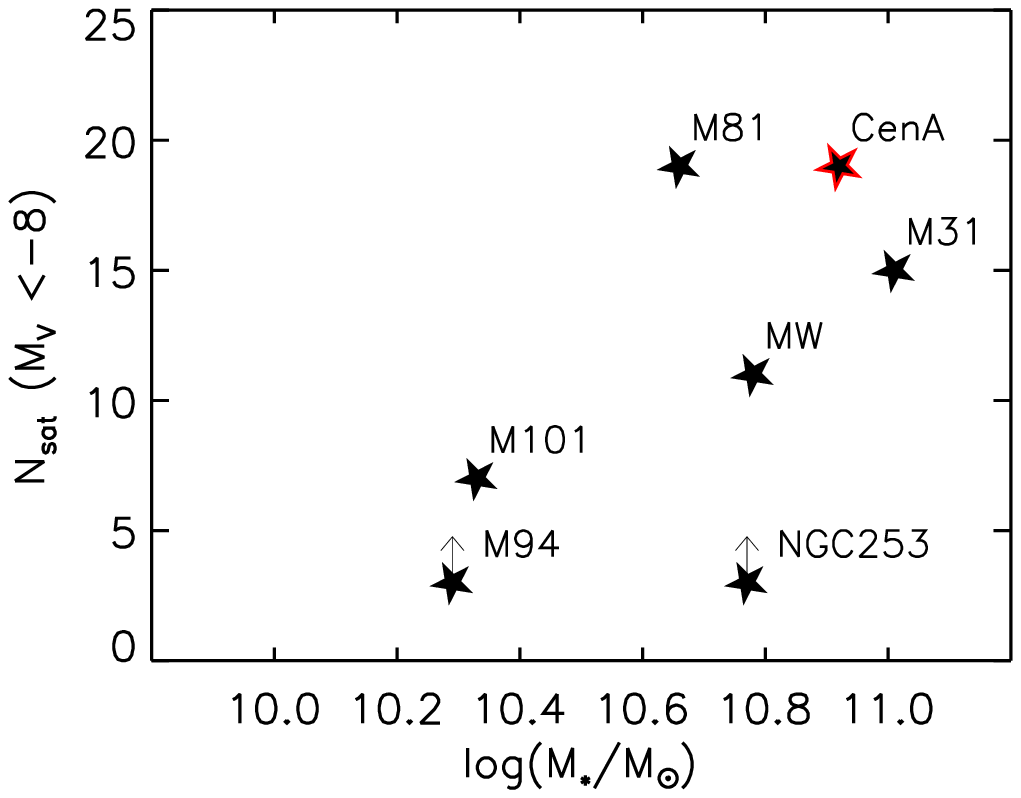}
\includegraphics[width = 0.475\linewidth]{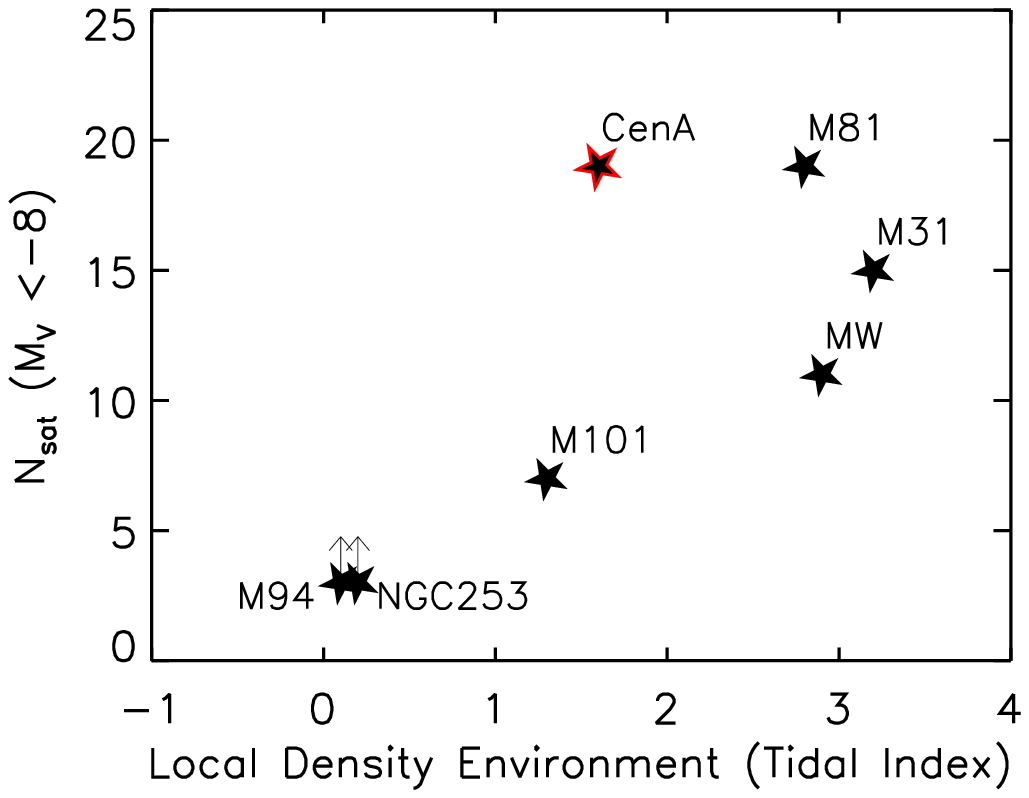}
\includegraphics[width = 0.475\linewidth]{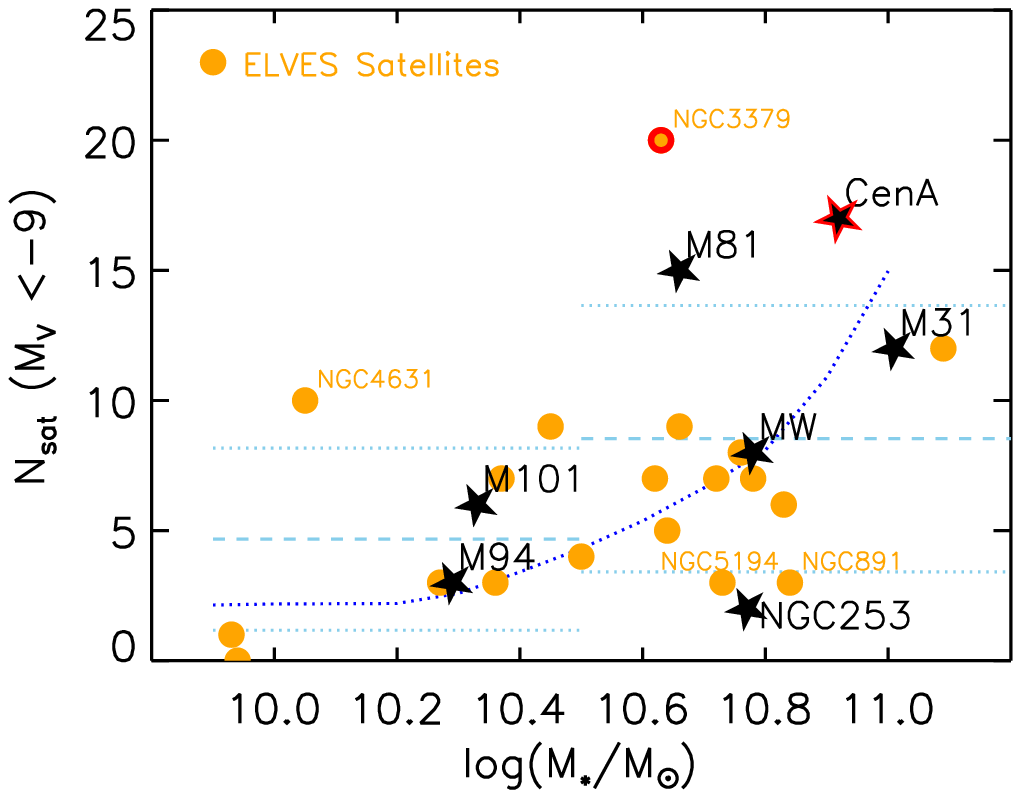}
\includegraphics[width = 0.475\linewidth]{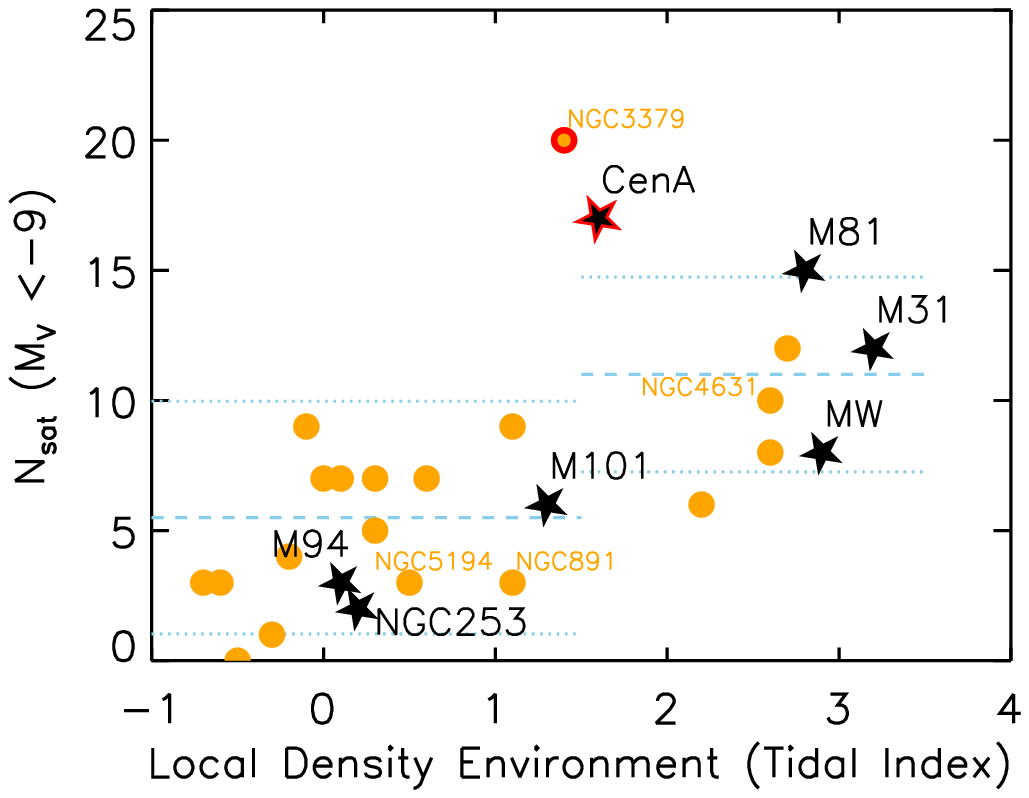}
\caption{Left panels: the relationship between the total satellite count and the host stellar mass. Right panels: the relationship between the total satellite count and local environment, based on tidal index ($\Theta_{5}$, where smaller numbers indicate a more isolated galaxy; see \citealt{Karachentsev2013}). The top panels include the satellites almost down to the ultra-faint regime ($M_V<-8$ within 150~kpc), and the bottom ones use the same magnitude cut as ELVES ($M_V<-9$ within 150~kpc) and include ELVES satellite statistics as a reference. The blue curved line in the bottom-left plot is taken from \citet{Danieli2023} and represents forward-modeled satellites with their fiducial stellar-to-halo mass relation based on ELVES satellites. The sample in the bottom panels is divided into two bins, based on host stellar mass ($\log(M_{\star}/M_{\odot})>$10.5 and $<10.5$, bottom-left) and tidal index ($\Theta_{5}>$1.5 and $<1.5$, bottom-right). The mean and standard deviation of each bin are depicted by dashed and dotted light blue lines, respectively. Elliptical galaxies (i.e., Cen~A and NGC~3379) are highlighted in red.
\label{fig:mass-tidal}}
\end{figure*}

Recently, \citet{Smercina2022} presented a tight linear relationship between the number of satellites and the largest merger partner ($M_{\star,Dom}$, the larger of either the current dominant satellite mass or the accreted mass), showing that systems with larger mergers host more satellites. The authors considered satellites within a projected radius of 150~kpc, down to $M_V<-9$. Figure~\ref{fig:smercina} shows this relationship for seven MW-mass systems for which we compiled a LF. Given that six out of these seven systems were also included in the \citeauthor{Smercina2022} sample, the linear trend in our sample is not surprising. Interestingly, NGC~253 stands out as a clear outlier in this relation with fewer satellites compared to other galaxies with its dominant merger mass. We use the value for NGC~253's accreted mass from Figure~1 of \cite{Smercina2022}, which was estimated by integrating the star-count-scaled projected 2D density profile in the range of 10-40~kpc and multiplying by a factor of three \citep[as in][]{Harmsen2017}. In terms of dominant merger mass, the MW is the most similar to NGC~253. But while the MW has seven satellites with $M_V<-9$, NGC~253 has only one (Scl-MM-dw2). \citet{Smercina2022} also considered M51, which seems to be a strong outlier in their $N_{sat}-M_{\star,Dom}$ relation (with only three satellites and $\log(M_{\star,Dom}/M_{\odot})=10.4\pm0.3$), but the authors argued that M51's satellite population is likely incomplete. However, the discrepancy for NGC~253 cannot be attributed to the survey completeness: our PISCeS program is complete down to $M_V < -8$ but limited to 100~kpc, however thanks to the ELVES Survey, the census of NGC~253 is complete down to $M_V < -9$ out to $\sim$300~kpc \citep{Carlsten2022}. It is clear that NGC~253's satellite population is a critical case study to better understand the formation of other MW-mass galaxies and their satellites. 

Another correlation that emerges from recent research is between the total satellite count and the host stellar mass, $N_{sat}-M_{\star}$. This relationship has been investigated in two separate samples: one considering the SAGA satellites ($M_r<-12.3$ within 150~kpc, \citealt{Mao2021}), and another considering the ELVES satellites ($M_V<-9$ within 150~kpc, \citealt{Carlsten21,Danieli2023}). Figure~\ref{fig:mass-tidal}-left panels show this relation for our seven compiled satellite systems. The top one includes the satellites almost down to the ultra-faint regime ($M_V<-8$ within 150~kpc), and the bottom one uses the same magnitude cut as ELVES ($M_V<-9$ within 150~kpc) and includes the ELVES satellites as a reference. Elliptical galaxies (i.e., Cen~A and NGC~3379) are highlighted in red. We adopt the host stellar masses from Table~1 of \citet{Carlsten2022}. The blue curved line in the bottom-left panel is taken from \citet{Danieli2023} and represents forward-modeled satellites with their fiducial stellar-to-halo mass relation based on the ELVES satellites. While it is apparent that a trend exists in general, there is a large spread in the number of satellites across all host mass bins. Upon dividing the sample with $M_V<-9$ into two bins based on host stellar mass, the average number of satellites is $4.7\pm3.5$ for $\log(M_{\star}/M_{\odot})<$10.5, and $8.5\pm5.1$ for $\log(M_{\star}/M_{\odot})\geq 10.5$, as represented by light blue lines\footnote{The choice of broad bins was made due to the uneven distribution of host numbers in narrower bins.}. NGC~4631 stands out with its higher number of satellites in the lower mass bin. The large scatter in the higher mass bin primarily results from distinct outliers such as NGC~253, NGC~5194, M81, and NGC~3379. Despite having comparable stellar masses, the number of satellites for these outliers varies significantly, ranging from just 2 to 20 satellites. This might suggest the presence of other important factors influencing the formation and evolution histories of these outliers. 

A similar trend also exists when the fainter satellites are considered (Figure~\ref{fig:mass-tidal}, top-left panel). Yet, NGC~253 again seems to have fewer satellites for its host stellar mass. Given that our PISCeS NGC253 footprint only extends out to 100~kpc, NGC~253 should be considered a lower limit in the top panels. However, it would be quite unusual for NGC~253 to host seven or more satellites located between 100 and 150~kpc. This discrepancy is also visible in the bottom-left panel with the brighter magnitude limit ($M_V<-9$), at which NGC~253's satellite census is complete. 

Figure~\ref{fig:mass-tidal}-right panels explore whether there is any correlation between satellite richness and environment, as proposed by \citet{Bennet19}, suggesting isolated MW-mass galaxies have fewer satellites than their counterparts in dense environments. The tidal index parameter (denoted as $\Theta_{5}$) was used as a measure of the galaxy environment, which takes into account the distance to and stellar mass of nearby galaxies \citep[see][]{Karachentsev2013}.
The ‘main disturber’ for a particular galaxy can be calculated via
\begin{equation}
\Theta_{1} = max[\log_{10}(M_{n}/D_{n}^{3})]+C, ~n=1, 2~ ... ~N
\end{equation}
\noindent where N is the total number of galaxies in the data set, $M_{n}$ is the mass of the neighboring galaxy, $D_{n}$ is the 3D separation between the galaxy and the neighboring galaxy, and C is a constant equal to -10.96, which has been chosen such that positive values of $\Theta_{1}$ indicate the membership in groups while negative values correspond to isolated galaxies. Because $\Theta_{1}$ can significantly change with time due to the orbital motions of galaxies, \citet{Karachentsev2013} advocates for $\Theta_{5}$ (which includes effects of more than one disturbing galaxy) as a more robust measure of the galaxy environment:
\begin{equation}
\Theta_{5} = \log_{10}(\sum_{n=1}^{5} M_{n}/D_{n}^{3})+C
\end{equation}
The tidal index $\Theta_{5}$ is the summation of the tidal force magnitude from the five neighbors of a galaxy where this magnitude is the highest. For this work, we draw tidal indices from the Updated Nearby Galaxy Catalog (UNGC, \citealt{Karachentsev2013}, \url{https://www.sao.ru/lv/lvgdb/}). We note that the UNGC is complete down to $M_B\sim -11$~mag, and the tidal force magnitude contribution from systems fainter than this limit is expected to be minimal, so the reported tidal index values should not significantly change with recently discovered Local Volume dwarfs.   

\begin{figure*}[!t]
\centering
\includegraphics[width = 0.475\linewidth]{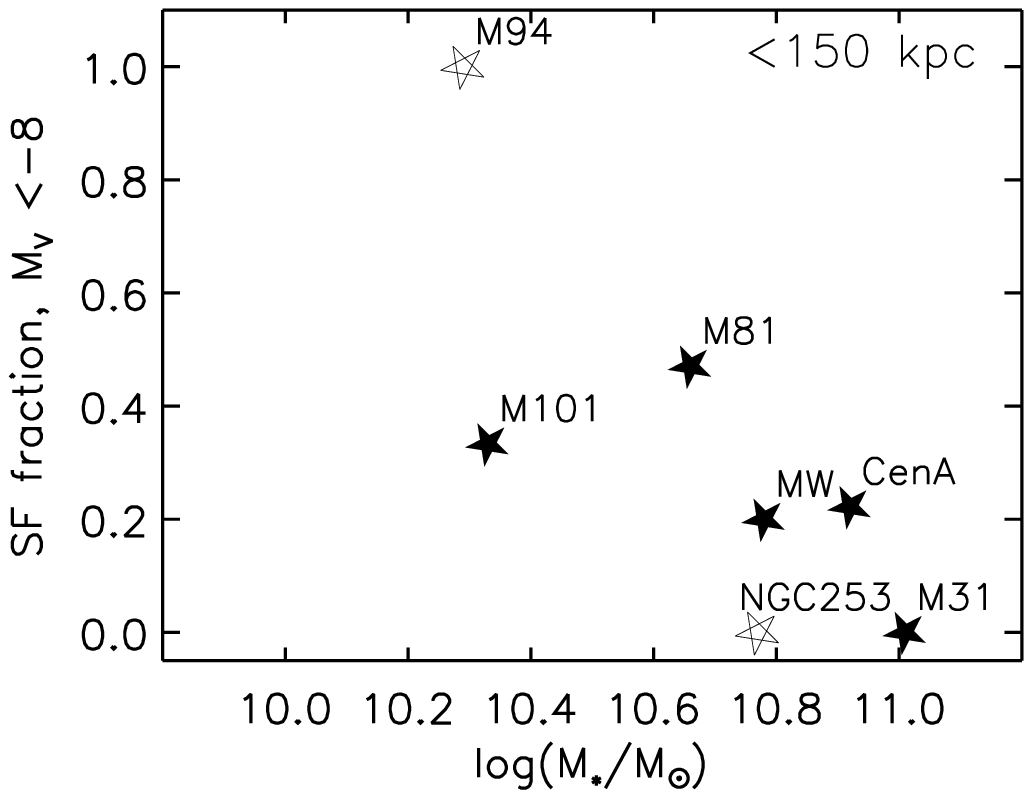}
\includegraphics[width = 0.475\linewidth]{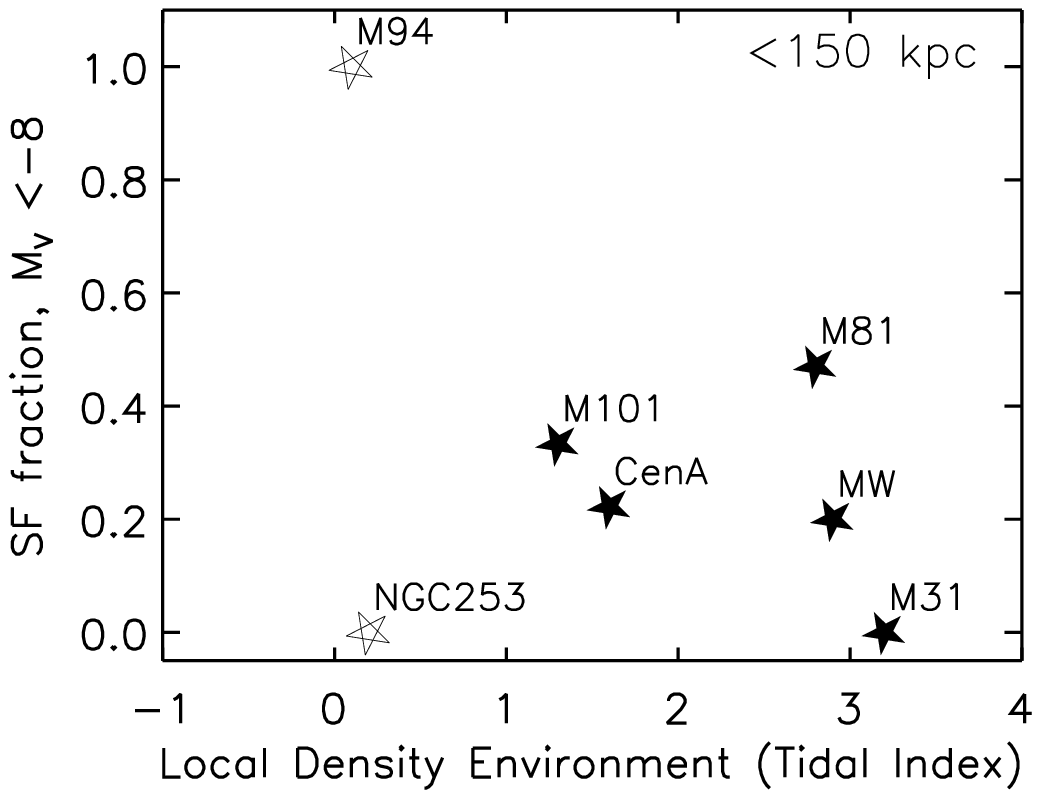}
\includegraphics[width = 0.475\linewidth]{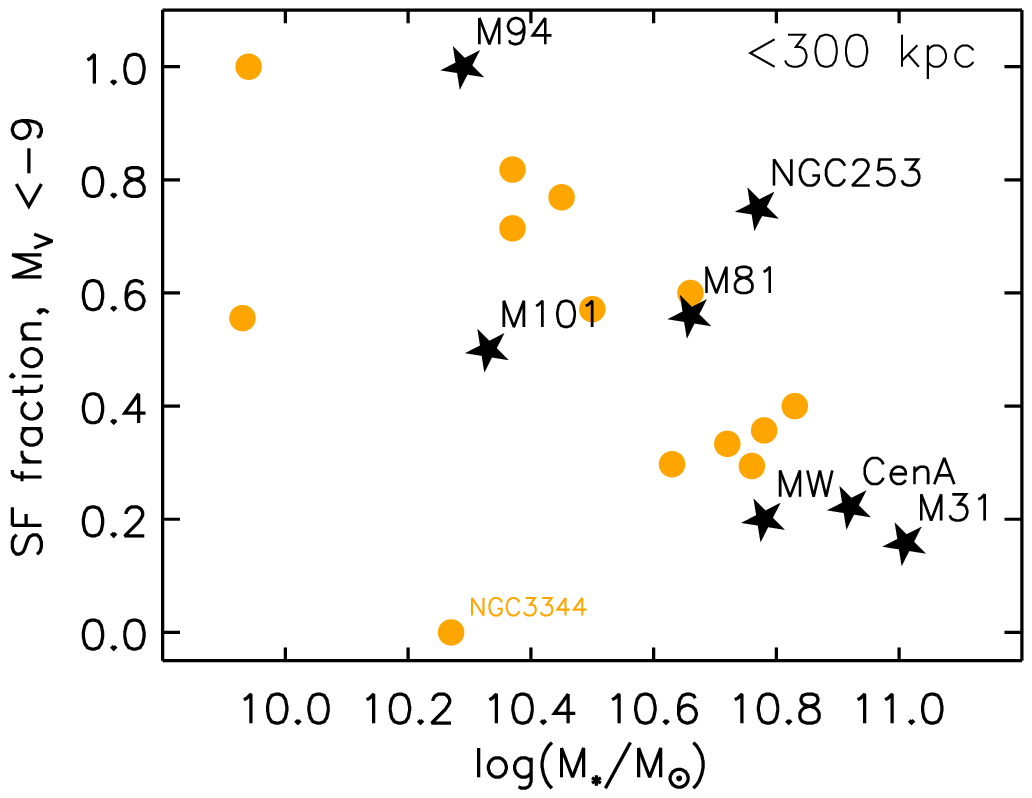}
\includegraphics[width = 0.475\linewidth]{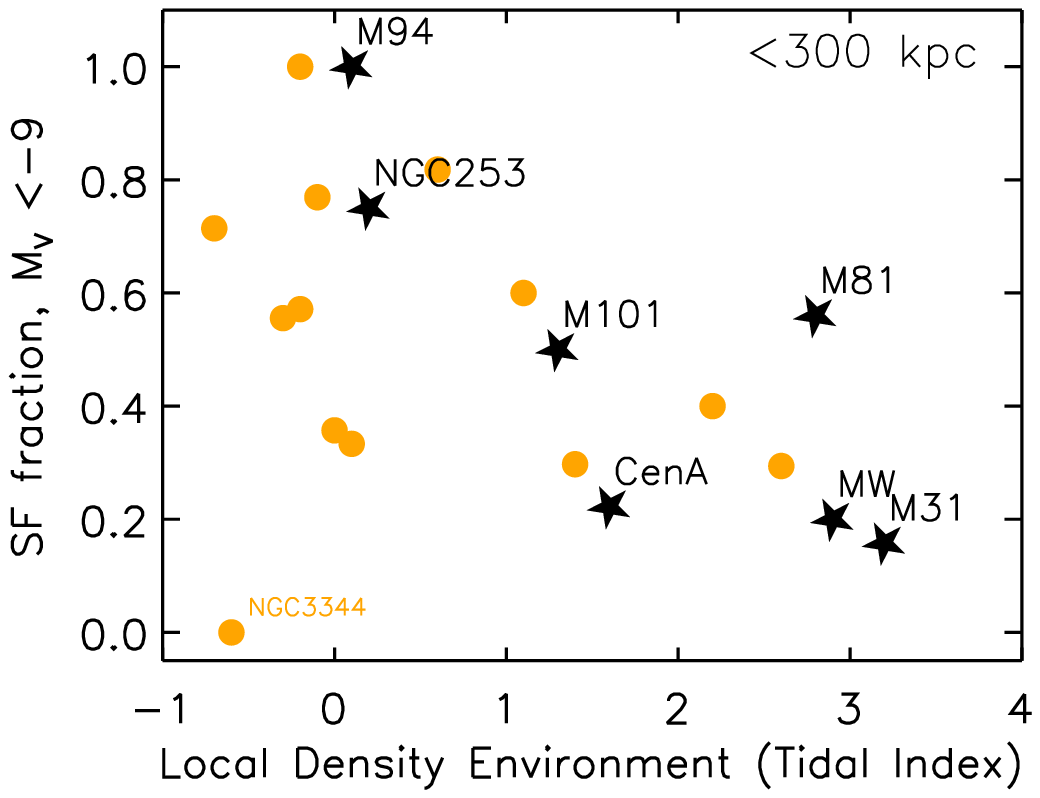}
\caption{The correlation between the star formation fraction and the host stellar mass is shown in the left panels, while the relationship with the tidal index is shown in the right panels. The top panels include the satellites almost down to the ultra-faint regime ($M_V<-8$ within 150~kpc). The bottom panels include the satellites with $M_V<-9$ within 300~kpc, along with ELVES satellites (orange circles). \label{fig:SF-trend}}
\end{figure*}

Figure~\ref{fig:mass-tidal}-right panels display the total satellite count as a function of the tidal index, and there is a clear trend where the objects with higher tidal indices also have more satellites. Upon dividing the sample with $M_V<-9$ into two bins\footnote{The choice of broad bins was made due to the uneven distribution of host numbers in narrower bins.} based on the tidal index, the average number of satellites is $5.5\pm4.5$ for $\Theta_{5}<$1.5, and $11.0\pm3.7$ for $\Theta_{5}\geq1.5$, as represented by light blue lines. Notably, NGC~253 no longer stands out with its fewer satellites, and neither does NGC~5194. Similarly, with their higher tidal indices, the satellite richness of both NGC~4631 and M81 agrees well with the trend. On the other hand, Cen~A and NGC~3379 are clear outliers with a relatively higher number of satellites for their tidal indices. Interestingly, they are the only elliptical galaxies in the sample plotted here, and recent works \citep[e.g.,][]{Javanmardi2020,Muller2023} found a correlation between the size of the bulge of a galaxy and the number of its dwarf galaxy satellites. For a given stellar mass of the host galaxy, their correlation is mainly driven by the morphology, where elliptical galaxies have more satellites than spiral galaxies. This morphology dependence could explain the larger number of satellites observed in Cen~A and NGC~3379. 

In short, host stellar mass, environment, and morphology all seem interconnected. All the relationships point to a complex picture of satellite formation and accretion where various properties appear in relation, and a successful model has to reproduce all of these trends. 

\subsection{Star-Forming Fractions}

There have been several recent studies investigating the environmental effects on the evolution of dwarf galaxies around MW-mass hosts (e.g., \citealt{Karunakaran21, Karunakaran22, Karunakaran2023, Greene2023, Jones2023, Christensen2023,Bhattacharyya2023}). Here we focus on the star-forming fraction of the satellite galaxies, and investigate if there are any trends as a function of host stellar mass and tidal index. We classify a system as star-forming if it shows any \hi~ detection, UV detection, or young stellar population in resolved stars ($\lesssim 1$~Gyr), and then calculate the star-forming fractions of our compiled satellite systems. Figure~\ref{fig:SF-trend} shows these fractions with respect to the host stellar mass in the left panels and the tidal index in the right panels. The top panels include satellites down to $M_V$$\approx -8$ within 150~kpc. We remind the reader that the M94 sample is complete only for magnitudes down to $M_V\approx-9$ and the NGC~253 sample is complete down to $M_V\approx-8$ out to 100~kpc. Therefore, both M94 and NGC~253 are shown with a hollow star to emphasize that their values are not final and might change with new faint dwarf discoveries.  The bottom panels use a brighter magnitude cut ($M_V<-9$) and include satellites out to 300~kpc, along with the ELVES satellites (orange circles) for comparison. Here we include ELVES hosts with a complete satellite census out to 300~kpc and those for which tidal index information is available. We adopt the ELVES star-forming fractions from \citet{Karunakaran2023}, which are primarily based on a combined UV detection and specific star formation rate criterion of confirmed satellites (see their Section~3 for more details). 

We observe an apparent trend with host stellar mass, where hosts with lower stellar mass have more star-forming satellites. It persists consistently whether we focus on the fainter inner satellites (top-left panel) or consider slightly brighter satellites out to 300~kpc (bottom-left panel). This trend is not unexpected, considering the correlation between stellar mass and halo mass. A higher stellar mass implies a larger halo, leading to a more substantial hot coronal gas reservoir and, consequently, increased ram pressure. Additionally, a more massive halo exerts a stronger gravitational influence on the satellite, resulting in enhanced tidal effects. If ram pressure and tides play an important role in quenching star formation, it logically follows that the star-forming fraction would vary with host stellar mass. 

Notably, NGC~253 stands out as an outlier with its high star-forming fraction for its stellar mass. However, this high fraction might be explained by its low-density environment (lower tidal index). While less pronounced, star-forming fractions appear to vary with tidal index, as proposed by \citet{Bennet19}: hosts in low-density environments have a higher star-forming fraction in satellites (see Figure~\ref{fig:SF-trend}). This trend is less evident when only inner satellites are considered (top-right panel), becoming more pronounced when satellites extending out to 300~kpc are included (bottom-right panel). There is a substantial scatter, particularly towards the lower tidal index range, which may be partially attributed to differences in host stellar mass. However, NGC~3344's satellite population deviates from these two trends: despite the expected abundance of star-forming satellites due to its relatively low stellar mass and low tidal index, all its confirmed satellites are observed to be quenched. It is worth noting that most ELVES satellites are confirmed via surface brightness fluctuation (SBF) distance measurements, which is not ideal for gas-rich, star-forming systems \citep{Greco2021}. This might introduce a potential bias in the sample, favoring quenched systems. Therefore, it is crucial to follow up on dwarf candidates in the ELVES sample, confirming and characterizing them to better understand the trends discussed here.  

\section{Conclusions}\label{sec:conclusion}
We have conducted a systematic search for resolved dwarf galaxies in the NGC~253 PISCeS data set. We statistically characterize our overall satellite detection efficiency in our survey, present {\it HST} follow-up of dwarf candidates beyond the survey footprint, investigate the existence of a satellite plane, and derive the dwarf galaxy luminosity function of NGC~253 to compare with those calculated for other Local Volume groups of galaxies. Here, we summarize our key results:

\begin{itemize}
    \item As a result of our systematic, complete search,  we recover three of the five NGC~253 PISCeS dwarfs, while the two others (Scl-MM-dw4 and Scl-MM-dw5) are UFDs that fall in the fields with the worst seeing. No new, high-confidence satellite galaxy candidates are discovered.
    \item We quantify the observational sensitivity of our search in terms of satellite properties (i.e., absolute magnitude, physical size), and find that our search is complete down to $M_V \sim -8$ and $\mu_V\sim$28~mag arcsec$^{-2}$, with a recovery rate of $>95$ \%: NGC~253 has only two classical satellites within 100~kpc, and the completeness for UFDs at $M_V \sim -7.5$ is about 30-40\%. Beyond our footprint, the census of NGC~253 satellites is complete down to $M_V \sim -9$ out to 300~kpc, based on ELVES \citep{Carlsten2022}. 
    \item We present deep {\it HST} follow-up observations of four dwarf candidates that were discovered beyond the PISCeS footprint: Do III, Do IV, and dw0036m2828 are confirmed to be satellites of NGC253, while ScuSR is found to be a background galaxy. We derive robust distances, luminosities, and structural parameters of Do~III, Do~IV, and dw0036m2828. All three are comparable to known Local Volume dwarf galaxies. While both Do~III and Do~IV have a handful of AGB stars consistent with a population of $\gtrsim6–8$~Gyr (similar to Scl-MM-dw1), dw0036m2828 has young stellar populations of ages between $100$ and $500$~Myr.
    \item We find no convincing evidence for the presence of a plane of satellites surrounding NGC~253, and argue that the asymmetric distribution of satellites in NGC~253 may be explained simply by the presence of an NGC~247 sub-group, as is expected in the $\Lambda$CDM paradigm.
    \item We compiled the LF of NGC~253 within 100 and 300~kpc. While the overall NGC~253 LF is consistent with the Local Volume sample, its slope is more similar with those of the relatively isolated M94 and M101 galaxies, suggesting a possible correlation with the surrounding environment.
    \item For both the $N_{sat}-M_{\star,Dom}$ and $N_{sat}-M_{\star}$ relationships, NGC~253 appears as an outlier with fewer satellites than expected. When considering the environment, as indicated by the tidal index, NGC~253 no longer stands out with its fewer satellites.  Instead, two elliptical galaxies in the Local Volume sample - Cen~A and NGC~3379 - emerge as outliers with a relatively higher number of satellites for their tidal indices. This observation aligns with recent research indicating a correlation between the size of a galaxy's bulge and the number of its dwarf galaxy satellites, where elliptical galaxies tend to have more satellites than spiral galaxies at a fixed host stellar mass \citep{Javanmardi2020,Muller2023}. Our work demonstrates that host stellar mass, environment, and morphology are all important players in dwarf satellite formation, highlighting the importance of exploring the faint end of the luminosity function of nearby galaxies across a spectrum of masses, morphologies, and environments.
    \item We focus on the star-forming fraction of satellite galaxies around MW-mass hosts, exploring trends related to host stellar mass and tidal index. We observe a consistent trend, indicating that hosts with lower stellar mass tend to have a higher proportion of star-forming satellites, observed in both fainter inner satellites and those extending to 300 kpc. NGC~253 stands out as an exception with a high star-forming fraction, potentially influenced by its low-density environment. While less prominent, a discernible trend with tidal index becomes notable, particularly when considering satellites extending to 300~kpc.
    
\end{itemize}

Our NGC~253 PISCeS survey has allowed us to study the faint end of the satellite function in a new, more isolated environment than the Local Group, extending almost down to the UFD regime. In the coming decade, it will be possible to go even further down the satellite luminosity function of NGC~253 and other systems in the Local Volume with the upcoming facilities like the Vera C. Rubin Observatory and its Legacy Survey of Space and Time (LSST; \citealt{Mutlupakdil21}). By employing well-optimized search approaches (such as machine learning–aided classification as in \citealt{Jones2023}), there is an opportunity to expand survey volumes effectively and establish a complete census of the faintest galaxies across {\it all} environments. The discovery of isolated faint dwarf galaxies is particularly intriguing, offering a unique reference sample for disentangling environmental galaxy formation processes from other mechanisms such as reionizaton and supernova feedback, as they live in fields isolated from galaxy groups where the environmental effect on galaxy processes is expected to be minimal \citep[e.g.,][]{Dickey2019}. Dwarf satellite research has a bright future with the upcoming surveys in the next decade, promising to illuminate many aspects of small-scale structure formation and galaxy evolution.

\vspace{5mm}
\facilities{Magellan: Clay (Megacam), Hubble Space Telescope, Green Bank Telescope, GALEX}

\software{Astropy \citep{astropy13,astropy18}, The IDL Astronomy User's Library \citep{IDLforever}, DOLPHOT \citep{Dolphin2000}}

\acknowledgements
Based on observations made with the NASA/ESA Hubble Space Telescope, obtained at the Space Telescope Science Institute, which is operated by the Association of Universities for Research in Astronomy, Inc., under NASA contract NAS5-26555. 
%All the {\it HST} data used in this paper can be found in MAST: XXX. 
These observations are associated with program \# HST-GO-17164.  Support for program \# HST-GO-17164 was provided by NASA through a grant from the Space Telescope Science Institute, which is operated by the Association of Universities for Research in Astronomy, Inc., under NASA contract NAS5-26555. This work was performed in part at the Aspen Center for Physics, which is supported by National Science Foundation grant PHY-2210452.
DJS acknowledges support from NSF grants AST-1821967, 1813708 and AST-2205863.

\appendix
\section{SculptorSR}\label{sec:appendix}
\begin{figure}[h]
\centering
\includegraphics[width = 0.475\linewidth]{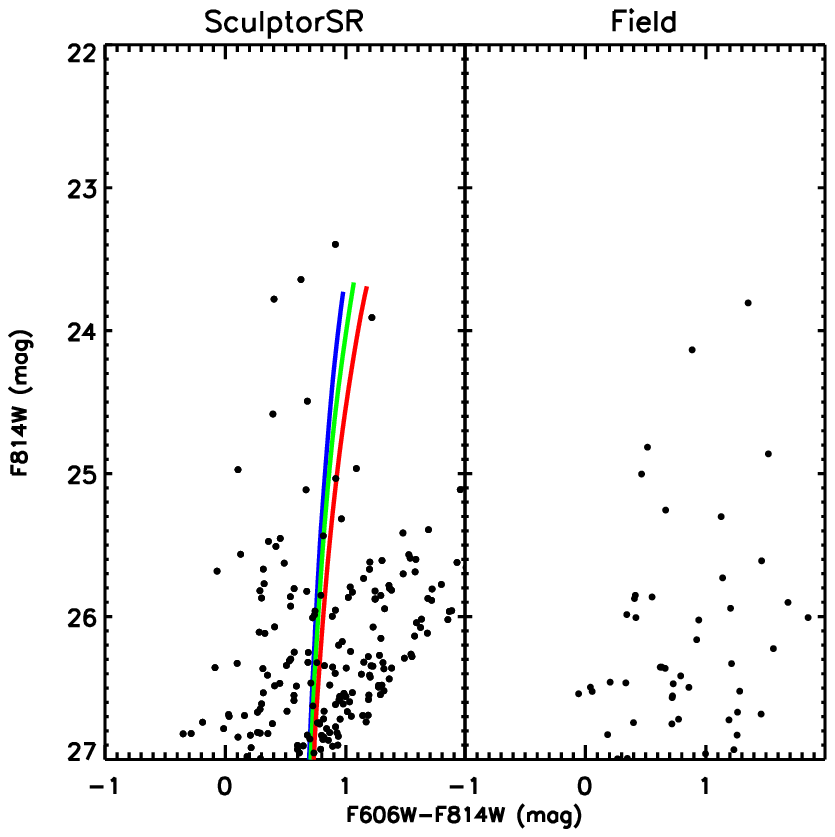}
\caption{Left: HST CMDs showing the stars within 0.5~arcmin of ScuSR. The blue, green, and red lines indicate the Dartmouth isochrones for 12 Gyr and [M/H] = -2.5 dex, -2.0 dex, and -1.5 dex, respectively. We shift each isochrone by the distance modulus of NGC~253. Right: CMD of a representative field region of equal area far away from ScuSR. Our follow-up HST observations show that ScuSR is a background galaxy, located well beyond NGC~253.
\label{fig:scsr}}
\end{figure}

Our follow-up {\it HST} observations show that ScuSR is a background galaxy, located well beyond NGC~253.  Figure~\ref{fig:scsr} shows its {\it HST} CMD displaying the stars within 0.5~arcmin, along with a CMD of a field region of equal size (see Section~\ref{sec:hst} for details of data and photometry). Isochrones are the same as in Figure~\ref{fig:cmd}, shifted to the distance of NGC~253. While ScuSR partially resolves into individual stars, those stars are not consistent with being at the distance of NGC~253 but are likely AGB stars of a background dwarf galaxy undergoing tidal interaction with NGC~150 (at $\sim$19~Mpc, \citealt{Springob2005}, see also Table~5 of \citealt{Carlsten2022}.) The current dataset is not sufficient to make detailed assessments of this galaxy. 

\bibliographystyle{aasjournal}
\bibliography{reference}

\begin{thebibliography}{}
\expandafter\ifx\csname natexlab\endcsname\relax\def\natexlab#1{#1}\fi
\providecommand{\url}[1]{\href{#1}{#1}}

\bibitem[{{Abbott} {et~al.}(2021){Abbott}, {Adam{\'o}w}, {Aguena}, {Allam}, {Amon}, {Annis}, {Avila}, {Bacon}, {Banerji}, {Bechtol}, {Becker}, {Bernstein}, {Bertin}, {Bhargava}, {Bridle}, {Brooks}, {Burke}, {Carnero Rosell}, {Carrasco Kind}, {Carretero}, {Castander}, {Cawthon}, {Chang}, {Choi}, {Conselice}, {Costanzi}, {Crocce}, {da Costa}, {Davis}, {De Vicente}, {DeRose}, {Desai}, {Diehl}, {Dietrich}, {Drlica-Wagner}, {Eckert}, {Elvin-Poole}, {Everett}, {Evrard}, {Ferrero}, {Fert{\'e}}, {Flaugher}, {Fosalba}, {Friedel}, {Frieman}, {Garc{\'\i}a-Bellido}, {Gaztanaga}, {Gelman}, {Gerdes}, {Giannantonio}, {Gill}, {Gruen}, {Gruendl}, {Gschwend}, {Gutierrez}, {Hartley}, {Hinton}, {Hollowood}, {Honscheid}, {Huterer}, {James}, {Jeltema}, {Johnson}, {Kent}, {Kron}, {Kuehn}, {Kuropatkin}, {Lahav}, {Li}, {Lidman}, {Lin}, {MacCrann}, {Maia}, {Manning}, {Maloney}, {March}, {Marshall}, {Martini}, {Melchior}, {Menanteau}, {Miquel}, {Morgan}, {Myles}, {Neilsen}, {Ogando}, {Palmese}, {Paz-Chinch{\'o}n}, {Petravick},
  {Pieres}, {Plazas}, {Pond}, {Rodriguez-Monroy}, {Romer}, {Roodman}, {Rykoff}, {Sako}, {Sanchez}, {Santiago}, {Scarpine}, {Serrano}, {Sevilla-Noarbe}, {Smith}, {Smith}, {Soares-Santos}, {Suchyta}, {Swanson}, {Tarle}, {Thomas}, {To}, {Tremblay}, {Troxel}, {Tucker}, {Turner}, {Varga}, {Walker}, {Wechsler}, {Weller}, {Wester}, {Wilkinson}, {Yanny}, {Zhang}, {Nikutta}, {Fitzpatrick}, {Jacques}, {Scott}, {Olsen}, {Huang}, {Herrera}, {Juneau}, {Nidever}, {Weaver}, {Adean}, {Correia}, {de Freitas}, {Freitas}, {Singulani}, {Vila-Verde}, \& {Linea Science Server}}]{DESDR2}
{Abbott}, T.~M.~C., {Adam{\'o}w}, M., {Aguena}, M., {et~al.} 2021, \apjs, 255, 20

\bibitem[{{Astropy Collaboration} {et~al.}(2013){Astropy Collaboration}, {Robitaille}, {Tollerud}, {Greenfield}, {Droettboom}, {Bray}, {Aldcroft}, {Davis}, {Ginsburg}, {Price-Whelan}, {Kerzendorf}, {Conley}, {Crighton}, {Barbary}, {Muna}, {Ferguson}, {Grollier}, {Parikh}, {Nair}, {Unther}, {Deil}, {Woillez}, {Conseil}, {Kramer}, {Turner}, {Singer}, {Fox}, {Weaver}, {Zabalza}, {Edwards}, {Azalee Bostroem}, {Burke}, {Casey}, {Crawford}, {Dencheva}, {Ely}, {Jenness}, {Labrie}, {Lim}, {Pierfederici}, {Pontzen}, {Ptak}, {Refsdal}, {Servillat}, \& {Streicher}}]{astropy13}
{Astropy Collaboration}, {Robitaille}, T.~P., {Tollerud}, E.~J., {et~al.} 2013, \aap, 558, A33

\bibitem[{{Astropy Collaboration} {et~al.}(2018){Astropy Collaboration}, {Price-Whelan}, {Sip{\H o}cz}, {G{\"u}nther}, {Lim}, {Crawford}, {Conseil}, {Shupe}, {Craig}, {Dencheva}, {Ginsburg}, {VanderPlas}, {Bradley}, {P{\'e}rez-Su{\'a}rez}, {de Val-Borro}, {Aldcroft}, {Cruz}, {Robitaille}, {Tollerud}, {Ardelean}, {Babej}, {Bach}, {Bachetti}, {Bakanov}, {Bamford}, {Barentsen}, {Barmby}, {Baumbach}, {Berry}, {Biscani}, {Boquien}, {Bostroem}, {Bouma}, {Brammer}, {Bray}, {Breytenbach}, {Buddelmeijer}, {Burke}, {Calderone}, {Cano Rodr{\'{\i}}guez}, {Cara}, {Cardoso}, {Cheedella}, {Copin}, {Corrales}, {Crichton}, {D'Avella}, {Deil}, {Depagne}, {Dietrich}, {Donath}, {Droettboom}, {Earl}, {Erben}, {Fabbro}, {Ferreira}, {Finethy}, {Fox}, {Garrison}, {Gibbons}, {Goldstein}, {Gommers}, {Greco}, {Greenfield}, {Groener}, {Grollier}, {Hagen}, {Hirst}, {Homeier}, {Horton}, {Hosseinzadeh}, {Hu}, {Hunkeler}, {Ivezi{\'c}}, {Jain}, {Jenness}, {Kanarek}, {Kendrew}, {Kern}, {Kerzendorf}, {Khvalko}, {King}, {Kirkby}, {Kulkarni},
  {Kumar}, {Lee}, {Lenz}, {Littlefair}, {Ma}, {Macleod}, {Mastropietro}, {McCully}, {Montagnac}, {Morris}, {Mueller}, {Mumford}, {Muna}, {Murphy}, {Nelson}, {Nguyen}, {Ninan}, {N{\"o}the}, {Ogaz}, {Oh}, {Parejko}, {Parley}, {Pascual}, {Patil}, {Patil}, {Plunkett}, {Prochaska}, {Rastogi}, {Reddy Janga}, {Sabater}, {Sakurikar}, {Seifert}, {Sherbert}, {Sherwood-Taylor}, {Shih}, {Sick}, {Silbiger}, {Singanamalla}, {Singer}, {Sladen}, {Sooley}, {Sornarajah}, {Streicher}, {Teuben}, {Thomas}, {Tremblay}, {Turner}, {Terr{\'o}n}, {van Kerkwijk}, {de la Vega}, {Watkins}, {Weaver}, {Whitmore}, {Woillez}, {Zabalza}, \& {Astropy Contributors}}]{astropy18}
{Astropy Collaboration}, {Price-Whelan}, A.~M., {Sip{\H o}cz}, B.~M., {et~al.} 2018, \aj, 156, 123

\bibitem[{{Bailin} {et~al.}(2011){Bailin}, {Bell}, {Chappell}, {Radburn-Smith}, \& {de Jong}}]{Bailin2011}
{Bailin}, J., {Bell}, E.~F., {Chappell}, S.~N., {Radburn-Smith}, D.~J., \& {de Jong}, R.~S. 2011, \apj, 736, 24

\bibitem[{{Bechtol} {et~al.}(2015){Bechtol}, {Drlica-Wagner}, {Balbinot}, {Pieres}, {Simon}, {Yanny}, {Santiago}, {Wechsler}, {Frieman}, {Walker}, {Williams}, {Rozo}, {Rykoff}, {Queiroz}, {Luque}, {Benoit-L{\'e}vy}, {Tucker}, {Sevilla}, {Gruendl}, {da Costa}, {Fausti Neto}, {Maia}, {Abbott}, {Allam}, {Armstrong}, {Bauer}, {Bernstein}, {Bernstein}, {Bertin}, {Brooks}, {Buckley-Geer}, {Burke}, {Carnero Rosell}, {Castander}, {Covarrubias}, {D'Andrea}, {DePoy}, {Desai}, {Diehl}, {Eifler}, {Estrada}, {Evrard}, {Fernandez}, {Finley}, {Flaugher}, {Gaztanaga}, {Gerdes}, {Girardi}, {Gladders}, {Gruen}, {Gutierrez}, {Hao}, {Honscheid}, {Jain}, {James}, {Kent}, {Kron}, {Kuehn}, {Kuropatkin}, {Lahav}, {Li}, {Lin}, {Makler}, {March}, {Marshall}, {Martini}, {Merritt}, {Miller}, {Miquel}, {Mohr}, {Neilsen}, {Nichol}, {Nord}, {Ogando}, {Peoples}, {Petravick}, {Plazas}, {Romer}, {Roodman}, {Sako}, {Sanchez}, {Scarpine}, {Schubnell}, {Smith}, {Soares-Santos}, {Sobreira}, {Suchyta}, {Swanson}, {Tarle}, {Thaler}, {Thomas},
  {Wester}, {Zuntz}, \& {DES Collaboration}}]{Bechtol2015}
{Bechtol}, K., {Drlica-Wagner}, A., {Balbinot}, E., {et~al.} 2015, \apj, 807, 50

\bibitem[{{Bell} {et~al.}(2022){Bell}, {Smercina}, {Price}, {D'Souza}, {Bailin}, {de Jong}, {Gozman}, {Jang}, {Monachesi}, {Gnedin}, \& {Slater}}]{Bell2022}
{Bell}, E.~F., {Smercina}, A., {Price}, P.~A., {et~al.} 2022, \apjl, 937, L3

\bibitem[{{Bennet} {et~al.}(2019){Bennet}, {Sand}, {Crnojevi{\'c}}, {Spekkens}, {Karunakaran}, {Zaritsky}, \& {Mutlu-Pakdil}}]{Bennet19}
{Bennet}, P., {Sand}, D.~J., {Crnojevi{\'c}}, D., {et~al.} 2019, \apj, 885, 153

\bibitem[{{Bennet} {et~al.}(2020){Bennet}, {Sand}, {Crnojevi{\'c}}, {Spekkens}, {Karunakaran}, {Zaritsky}, \& {Mutlu-Pakdil}}]{Bennet20}
---. 2020, \apjl, 893, L9

\bibitem[{{Bennet} {et~al.}(2017){Bennet}, {Sand}, {Crnojevi{\'c}}, {Spekkens}, {Zaritsky}, \& {Karunakaran}}]{Bennet17}
---. 2017, \apj, 850, 109

\bibitem[{{Bhattacharyya} {et~al.}(2023){Bhattacharyya}, {Peter}, {Martini}, {Mutlu-Pakdil}, {Drlica-Wagner}, {Pace}, {Strigari}, {Cheng}, {Roberts}, {Tanoglidis}, {Aguena}, {Alves}, {Andrade-Oliveira}, {Bacon}, {Brooks}, {Carnero Rosell}, {Carretero}, {da Costa}, {Pereira}, {Davis}, {Desai}, {Doel}, {Ferrero}, {Frieman}, {Garc{\'\i}a-Bellido}, {Giannini}, {Gruen}, {Gruendl}, {Hinton}, {Hollowood}, {Honscheid}, {James}, {Kuehn}, {Marshall}, {Mena-Fern{\'a}ndez}, {Miquel}, {Palmese}, {Pieres}, {Plazas Malag{\'o}n}, {Sanchez}, {Santiago}, {Schubnell}, {Sevilla-Noarbe}, {Smith}, {Suchyta}, {Swanson}, {Tarle}, {Vincenzi}, {Walker}, {Weaverdyck}, \& {Wiseman}}]{Bhattacharyya2023}
{Bhattacharyya}, J., {Peter}, A.~H.~G., {Martini}, P., {et~al.} 2023, arXiv e-prints, arXiv:2312.00773

\bibitem[{{Bouchard} {et~al.}(2005){Bouchard}, {Jerjen}, {Da Costa}, \& {Ott}}]{Bouchard2005}
{Bouchard}, A., {Jerjen}, H., {Da Costa}, G.~S., \& {Ott}, J. 2005, \aj, 130, 2058

\bibitem[{{Bressan} {et~al.}(2012){Bressan}, {Marigo}, {Girardi}, {Salasnich}, {Dal Cero}, {Rubele}, \& {Nanni}}]{Bressan2012}
{Bressan}, A., {Marigo}, P., {Girardi}, L., {et~al.} 2012, \mnras, 427, 127

\bibitem[{{Calzetti}(2013)}]{Calzetti_2013}
{Calzetti}, D. 2013, in Secular Evolution of Galaxies, ed. J.~{Falc{\'o}n-Barroso} \& J.~H. {Knapen}, 419

\bibitem[{{Cannon} {et~al.}(2011){Cannon}, {Giovanelli}, {Haynes}, {Janowiecki}, {Parker}, {Salzer}, {Adams}, {Engstrom}, {Huang}, {McQuinn}, {Ott}, {Saintonge}, {Skillman}, {Allan}, {Erny}, {Fliss}, \& {Smith}}]{Cannon2011}
{Cannon}, J.~M., {Giovanelli}, R., {Haynes}, M.~P., {et~al.} 2011, \apjl, 739, L22

\bibitem[{{Carlin} {et~al.}(2016){Carlin}, {Sand}, {Price}, {Willman}, {Karunakaran}, {Spekkens}, {Bell}, {Brodie}, {Crnojevi{\'c}}, {Forbes}, {Hargis}, {Kirby}, {Lupton}, {Peter}, {Romanowsky}, \& {Strader}}]{Carlin16}
{Carlin}, J.~L., {Sand}, D.~J., {Price}, P., {et~al.} 2016, \apjl, 828, L5

\bibitem[{{Carlin} {et~al.}(2021){Carlin}, {Mutlu-Pakdil}, {Crnojevi{\'c}}, {Garling}, {Karunakaran}, {Peter}, {Tollerud}, {Forbes}, {Hargis}, {Lim}, {Romanowsky}, {Sand}, {Spekkens}, \& {Strader}}]{carlin21}
{Carlin}, J.~L., {Mutlu-Pakdil}, B., {Crnojevi{\'c}}, D., {et~al.} 2021, \apj, 909, 211

\bibitem[{{Carlsten} {et~al.}(2022){Carlsten}, {Greene}, {Beaton}, {Danieli}, \& {Greco}}]{Carlsten2022}
{Carlsten}, S.~G., {Greene}, J.~E., {Beaton}, R.~L., {Danieli}, S., \& {Greco}, J.~P. 2022, \apj, 933, 47

\bibitem[{{Carlsten} {et~al.}(2021){Carlsten}, {Greene}, {Greco}, {Beaton}, \& {Kado-Fong}}]{Carlsten21}
{Carlsten}, S.~G., {Greene}, J.~E., {Greco}, J.~P., {Beaton}, R.~L., \& {Kado-Fong}, E. 2021, arXiv e-prints, arXiv:2105.03435

\bibitem[{{Chiboucas} {et~al.}(2013){Chiboucas}, {Jacobs}, {Tully}, \& {Karachentsev}}]{Chiboucas13}
{Chiboucas}, K., {Jacobs}, B.~A., {Tully}, R.~B., \& {Karachentsev}, I.~D. 2013, \aj, 146, 126

\bibitem[{{Chiboucas} {et~al.}(2009){Chiboucas}, {Karachentsev}, \& {Tully}}]{Chiboucas2009}
{Chiboucas}, K., {Karachentsev}, I.~D., \& {Tully}, R.~B. 2009, \aj, 137, 3009

\bibitem[{{Christensen} {et~al.}(2023){Christensen}, {Brooks}, {Munshi}, {Riggs}, {Van Nest}, {Akins}, {Quinn}, \& {Chamberland}}]{Christensen2023}
{Christensen}, C.~R., {Brooks}, A., {Munshi}, F., {et~al.} 2023, arXiv e-prints, arXiv:2311.04975

\bibitem[{{Collins} {et~al.}(2022){Collins}, {Charles}, {Mart{\'\i}nez-Delgado}, {Monelli}, {Karim}, {Donatiello}, {Tollerud}, \& {Boschin}}]{Collins2022}
{Collins}, M. L.~M., {Charles}, E. J.~E., {Mart{\'\i}nez-Delgado}, D., {et~al.} 2022, \mnras, 515, L72

\bibitem[{{Crnojevi{\'c}} {et~al.}(2016){Crnojevi{\'c}}, {Sand}, {Spekkens}, {Caldwell}, {Guhathakurta}, {McLeod}, {Seth}, {Simon}, {Strader}, \& {Toloba}}]{Crnojevic16}
{Crnojevi{\'c}}, D., {Sand}, D.~J., {Spekkens}, K., {et~al.} 2016, \apj, 823, 19

\bibitem[{{Crnojevi{\'c}} {et~al.}(2019){Crnojevi{\'c}}, {Sand}, {Bennet}, {Pasetto}, {Spekkens}, {Caldwell}, {Guhathakurta}, {McLeod}, {Seth}, {Simon}, {Strader}, \& {Toloba}}]{Crnojevic19}
{Crnojevi{\'c}}, D., {Sand}, D.~J., {Bennet}, P., {et~al.} 2019, \apj, 872, 80

\bibitem[{{Crosby} {et~al.}(2023){Crosby}, {Jerjen}, {M{\"u}ller}, {Pawlowski}, {Mateo}, \& {Lelli}}]{Crosby2023}
{Crosby}, E., {Jerjen}, H., {M{\"u}ller}, O., {et~al.} 2023, \mnras, arXiv:2312.03486

\bibitem[{{Danieli} {et~al.}(2023){Danieli}, {Greene}, {Carlsten}, {Jiang}, {Beaton}, \& {Goulding}}]{Danieli2023}
{Danieli}, S., {Greene}, J.~E., {Carlsten}, S., {et~al.} 2023, \apj, 956, 6

\bibitem[{{Danieli} {et~al.}(2017){Danieli}, {van Dokkum}, {Merritt}, {Abraham}, {Zhang}, {Karachentsev}, \& {Makarova}}]{Danieli17}
{Danieli}, S., {van Dokkum}, P., {Merritt}, A., {et~al.} 2017, \apj, 837, 136

\bibitem[{{Davis} {et~al.}(2021){Davis}, {Nierenberg}, {Peter}, {Garling}, {Greco}, {Kochanek}, {Utomo}, {Casey}, {Pogge}, {Roberts}, {Sand}, \& {Sardone}}]{Davis21}
{Davis}, A.~B., {Nierenberg}, A.~M., {Peter}, A. H.~G., {et~al.} 2021, \mnras, 500, 3854

\bibitem[{{Dickey} {et~al.}(2019){Dickey}, {Geha}, {Wetzel}, \& {El-Badry}}]{Dickey2019}
{Dickey}, C.~M., {Geha}, M., {Wetzel}, A., \& {El-Badry}, K. 2019, \apj, 884, 180

\bibitem[{{Doliva-Dolinsky} {et~al.}(2022){Doliva-Dolinsky}, {Martin}, {Thomas}, {Ferguson}, {Ibata}, {Lewis}, {Mackey}, {McConnachie}, \& {Yuan}}]{Amandine2022}
{Doliva-Dolinsky}, A., {Martin}, N.~F., {Thomas}, G.~F., {et~al.} 2022, \apj, 933, 135

\bibitem[{{Dolphin}(2000)}]{Dolphin2000}
{Dolphin}, A.~E. 2000, \pasp, 112, 1383

\bibitem[{{Dotter} {et~al.}(2008){Dotter}, {Chaboyer}, {Jevremovi{\'c}}, {Kostov}, {Baron}, \& {Ferguson}}]{Dotter2008}
{Dotter}, A., {Chaboyer}, B., {Jevremovi{\'c}}, D., {et~al.} 2008, \apjs, 178, 89

\bibitem[{{Drlica-Wagner} {et~al.}(2015){Drlica-Wagner}, {Bechtol}, {Rykoff}, {Luque}, {Queiroz}, {Mao}, {Wechsler}, {Simon}, {Santiago}, {Yanny}, {Balbinot}, {Dodelson}, {Fausti Neto}, {James}, {Li}, {Maia}, {Marshall}, {Pieres}, {Stringer}, {Walker}, {Abbott}, {Abdalla}, {Allam}, {Benoit-L{\'e}vy}, {Bernstein}, {Bertin}, {Brooks}, {Buckley-Geer}, {Burke}, {Carnero Rosell}, {Carrasco Kind}, {Carretero}, {Crocce}, {da Costa}, {Desai}, {Diehl}, {Dietrich}, {Doel}, {Eifler}, {Evrard}, {Finley}, {Flaugher}, {Fosalba}, {Frieman}, {Gaztanaga}, {Gerdes}, {Gruen}, {Gruendl}, {Gutierrez}, {Honscheid}, {Kuehn}, {Kuropatkin}, {Lahav}, {Martini}, {Miquel}, {Nord}, {Ogando}, {Plazas}, {Reil}, {Roodman}, {Sako}, {Sanchez}, {Scarpine}, {Schubnell}, {Sevilla-Noarbe}, {Smith}, {Soares-Santos}, {Sobreira}, {Suchyta}, {Swanson}, {Tarle}, {Tucker}, {Vikram}, {Wester}, {Zhang}, {Zuntz}, \& {DES Collaboration}}]{Drlica-Wagner2015}
{Drlica-Wagner}, A., {Bechtol}, K., {Rykoff}, E.~S., {et~al.} 2015, \apj, 813, 109

\bibitem[{{Drlica-Wagner} {et~al.}(2021){Drlica-Wagner}, {Carlin}, {Nidever}, {Ferguson}, {Kuropatkin}, {Adam{\'o}w}, {Cerny}, {Choi}, {Esteves}, {Mart{\'\i}nez-V{\'a}zquez}, {Mau}, {Miller}, {Mutlu-Pakdil}, {Neilsen}, {Olsen}, {Pace}, {Riley}, {Sakowska}, {Sand}, {Santana-Silva}, {Tollerud}, {Tucker}, {Vivas}, {Zaborowski}, {Zenteno}, {Abbott}, {Allam}, {Bechtol}, {Bell}, {Bell}, {Bilaji}, {Bom}, {Carballo-Bello}, {Cioni}, {Diaz-Ocampo}, {de Boer}, {Erkal}, {Gruendl}, {Hernandez-Lang}, {Hughes}, {James}, {Johnson}, {Li}, {Mao}, {Mart{\'\i}nez-Delgado}, {Massana}, {McNanna}, {Morgan}, {Nadler}, {No{\"e}l}, {Palmese}, {Peter}, {Rykoff}, {S{\'a}nchez}, {Shipp}, {Simon}, {Smercina}, {Soares-Santos}, {Stringfellow}, {Tavangar}, {van der Marel}, {Walker}, {Wechsler}, {Wu}, {Yanny}, {Fitzpatrick}, {Huang}, {Jacques}, {Nikutta}, \& {Scott}}]{Drlica-Wagner2021}
{Drlica-Wagner}, A., {Carlin}, J.~L., {Nidever}, D.~L., {et~al.} 2021, arXiv e-prints, arXiv:2103.07476

\bibitem[{{Fan} {et~al.}(2023){Fan}, {Moon}, {Park}, {Zaritsky}, {Kim}, {Lee}, {Li}, {Ni}, {Shin}, {Cha}, \& {Lee}}]{Fan2023}
{Fan}, T.~J., {Moon}, D.-S., {Park}, H.~S., {et~al.} 2023, \mnras, 525, 4904

\bibitem[{{Geha} {et~al.}(2017){Geha}, {Wechsler}, {Mao}, {Tollerud}, {Weiner}, {Bernstein}, {Hoyle}, {Marchi}, {Marshall}, {Mu{\~n}oz}, \& {Lu}}]{Geha17}
{Geha}, M., {Wechsler}, R.~H., {Mao}, Y.-Y., {et~al.} 2017, \apj, 847, 4

\bibitem[{{Giovanelli} {et~al.}(2013){Giovanelli}, {Haynes}, {Adams}, {Cannon}, {Rhode}, {Salzer}, {Skillman}, {Bernstein-Cooper}, \& {McQuinn}}]{Giovanelli13}
{Giovanelli}, R., {Haynes}, M.~P., {Adams}, E. A.~K., {et~al.} 2013, \aj, 146, 15

\bibitem[{{Goto} {et~al.}(2023){Goto}, {Zaritsky}, {Karunakaran}, {Donnerstein}, \& {Sand}}]{Goto2023}
{Goto}, H., {Zaritsky}, D., {Karunakaran}, A., {Donnerstein}, R., \& {Sand}, D.~J. 2023, \aj, 166, 185

\bibitem[{{Greco} {et~al.}(2021){Greco}, {van Dokkum}, {Danieli}, {Carlsten}, \& {Conroy}}]{Greco2021}
{Greco}, J.~P., {van Dokkum}, P., {Danieli}, S., {Carlsten}, S.~G., \& {Conroy}, C. 2021, \apj, 908, 24

\bibitem[{{Greene} {et~al.}(2023){Greene}, {Danieli}, {Carlsten}, {Beaton}, {Jiang}, \& {Li}}]{Greene2023}
{Greene}, J.~E., {Danieli}, S., {Carlsten}, S., {et~al.} 2023, \apj, 949, 94

\bibitem[{{Harmsen} {et~al.}(2017){Harmsen}, {Monachesi}, {Bell}, {de Jong}, {Bailin}, {Radburn-Smith}, \& {Holwerda}}]{Harmsen2017}
{Harmsen}, B., {Monachesi}, A., {Bell}, E.~F., {et~al.} 2017, \mnras, 466, 1491

\bibitem[{{Harris} {et~al.}(2010){Harris}, {Rejkuba}, \& {Harris}}]{Harris2010}
{Harris}, G. L.~H., {Rejkuba}, M., \& {Harris}, W.~E. 2010, \pasa, 27, 457

\bibitem[{{Iglesias-P{\'a}ramo} {et~al.}(2006){Iglesias-P{\'a}ramo}, {Buat}, {Takeuchi}, {Xu}, {Boissier}, {Boselli}, {Burgarella}, {Madore}, {Gil de Paz}, {Bianchi}, {Barlow}, {Byun}, {Donas}, {Forster}, {Friedman}, {Heckman}, {Jelinski}, {Lee}, {Malina}, {Martin}, {Milliard}, {Morrissey}, {Neff}, {Rich}, {Schiminovich}, {Seibert}, {Siegmund}, {Small}, {Szalay}, {Welsh}, \& {Wyder}}]{IglesiasParamo2006}
{Iglesias-P{\'a}ramo}, J., {Buat}, V., {Takeuchi}, T.~T., {et~al.} 2006, \apjs, 164, 38

\bibitem[{{Jacobs} {et~al.}(2009){Jacobs}, {Rizzi}, {Tully}, {Shaya}, {Makarov}, \& {Makarova}}]{Jacobs2009}
{Jacobs}, B.~A., {Rizzi}, L., {Tully}, R.~B., {et~al.} 2009, \aj, 138, 332

\bibitem[{{Jang} \& {Lee}(2017)}]{jang17}
{Jang}, I.~S., \& {Lee}, M.~G. 2017, \apj, 835, 28

\bibitem[{{Javanmardi} \& {Kroupa}(2020)}]{Javanmardi2020}
{Javanmardi}, B., \& {Kroupa}, P. 2020, \mnras, 493, L44

\bibitem[{{Jerjen} {et~al.}(1998){Jerjen}, {Freeman}, \& {Binggeli}}]{Jerjen1998}
{Jerjen}, H., {Freeman}, K.~C., \& {Binggeli}, B. 1998, \aj, 116, 2873

\bibitem[{{Jones} {et~al.}(2023{\natexlab{a}}){Jones}, {Mutlu-Pakdil}, {Sand}, {Donnerstein}, {Crnojevi{\'c}}, {Bennet}, {Fielder}, {Karunakaran}, {Spekkens}, {Strader}, {Urquhart}, \& {Zaritsky}}]{Jones2023}
{Jones}, M.~G., {Mutlu-Pakdil}, B., {Sand}, D.~J., {et~al.} 2023{\natexlab{a}}, \apjl, 957, L5

\bibitem[{{Jones} {et~al.}(2023{\natexlab{b}}){Jones}, {Sand}, {Karunakaran}, {Spekkens}, {Oman}, {Bennet}, {Besla}, {Crnojevic}, {Cuillandre}, {Fielder}, {Gwyn}, \& {Mutlu-Pakdil}}]{Jones2023b}
{Jones}, M.~G., {Sand}, D.~J., {Karunakaran}, A., {et~al.} 2023{\natexlab{b}}, arXiv e-prints, arXiv:2311.02152

\bibitem[{{Kalberla} \& {Haud}(2015)}]{Kalberla15}
{Kalberla}, P.~M.~W., \& {Haud}, U. 2015, \aap, 578, A78

\bibitem[{{Kalberla} {et~al.}(2010){Kalberla}, {McClure-Griffiths}, {Pisano}, {Calabretta}, {Ford}, {Lockman}, {Staveley-Smith}, {Kerp}, {Winkel}, {Murphy}, \& {Newton-McGee}}]{Kalberla10}
{Kalberla}, P.~M.~W., {McClure-Griffiths}, N.~M., {Pisano}, D.~J., {et~al.} 2010, \aap, 521, A17

\bibitem[{{Karachentsev} {et~al.}(2013){Karachentsev}, {Makarov}, \& {Kaisina}}]{Karachentsev2013}
{Karachentsev}, I.~D., {Makarov}, D.~I., \& {Kaisina}, E.~I. 2013, \aj, 145, 101

\bibitem[{{Karachentsev} {et~al.}(2021){Karachentsev}, {Tully}, {Anand}, {Rizzi}, \& {Shaya}}]{Karachentsev2021}
{Karachentsev}, I.~D., {Tully}, R.~B., {Anand}, G.~S., {Rizzi}, L., \& {Shaya}, E.~J. 2021, \aj, 161, 205

\bibitem[{{Karachentsev} {et~al.}(2003){Karachentsev}, {Grebel}, {Sharina}, {Dolphin}, {Geisler}, {Guhathakurta}, {Hodge}, {Karachentseva}, {Sarajedini}, \& {Seitzer}}]{Karachentsev2003}
{Karachentsev}, I.~D., {Grebel}, E.~K., {Sharina}, M.~E., {et~al.} 2003, \aap, 404, 93

\bibitem[{{Karunakaran} {et~al.}(2023){Karunakaran}, {Sand}, {Jones}, {Spekkens}, {Bennet}, {Crnojevi{\'c}}, {Mutlu-Pakd{\i}{\ensuremath{\dot{}}}l}, \& {Zaritsky}}]{Karunakaran2023}
{Karunakaran}, A., {Sand}, D.~J., {Jones}, M.~G., {et~al.} 2023, \mnras, 524, 5314

\bibitem[{{Karunakaran} {et~al.}(2022){Karunakaran}, {Spekkens}, {Carroll}, {Sand}, {Bennet}, {Crnojevi{\'c}}, {Jones}, \& {Mutlu-Pakd{\i}l}}]{Karunakaran22}
{Karunakaran}, A., {Spekkens}, K., {Carroll}, R., {et~al.} 2022, \mnras, 516, 1741

\bibitem[{{Karunakaran} {et~al.}(2021){Karunakaran}, {Spekkens}, {Oman}, {Simpson}, {Fattahi}, {Sand}, {Bennet}, {Crnojevi{\'c}}, {Frenk}, {G{\'o}mez}, {Grand}, {Jones}, {Marinacci}, {Mutlu-Pakdil}, {Navarro}, \& {Zaritsky}}]{Karunakaran21}
{Karunakaran}, A., {Spekkens}, K., {Oman}, K.~A., {et~al.} 2021, arXiv e-prints, arXiv:2105.09321

\bibitem[{{Koposov} {et~al.}(2015){Koposov}, {Belokurov}, {Torrealba}, \& {Evans}}]{Koposov2015}
{Koposov}, S.~E., {Belokurov}, V., {Torrealba}, G., \& {Evans}, N.~W. 2015, \apj, 805, 130

\bibitem[{{Kroupa}(2001)}]{Kroupa2001}
{Kroupa}, P. 2001, \mnras, 322, 231

\bibitem[{{Landsman}(1993)}]{IDLforever}
{Landsman}, W.~B. 1993, in Astronomical Society of the Pacific Conference Series, Vol.~52, Astronomical Data Analysis Software and Systems II, ed. R.~J. {Hanisch}, R.~J.~V. {Brissenden}, \& J.~{Barnes}, 246

\bibitem[{{Lee} {et~al.}(1993){Lee}, {Freedman}, \& {Madore}}]{Lee1993}
{Lee}, M.~G., {Freedman}, W.~L., \& {Madore}, B.~F. 1993, \apj, 417, 553

\bibitem[{{Longeard} {et~al.}(2021){Longeard}, {Martin}, {Ibata}, {Starkenburg}, {Jablonka}, {Aguado}, {Carlberg}, {C{\^o}t{\'e}}, {Gonz{\'a}lez Hern{\'a}ndez}, {Lucchesi}, {Malhan}, {Navarro}, {S{\'a}nchez-Janssen}, {Thomas}, {Venn}, \& {McConnachie}}]{Longeard2021}
{Longeard}, N., {Martin}, N., {Ibata}, R.~A., {et~al.} 2021, \mnras, 503, 2754

\bibitem[{{Mao} {et~al.}(2021){Mao}, {Geha}, {Wechsler}, {Weiner}, {Tollerud}, {Nadler}, \& {Kallivayalil}}]{Mao2021}
{Mao}, Y.-Y., {Geha}, M., {Wechsler}, R.~H., {et~al.} 2021, \apj, 907, 85

\bibitem[{{Martin} \& {GALEX Team}(2005)}]{galex2005}
{Martin}, C., \& {GALEX Team}. 2005, in Maps of the Cosmos, ed. M.~{Colless}, L.~{Staveley-Smith}, \& R.~A. {Stathakis}, Vol. 216, 221

\bibitem[{{Martin} {et~al.}(2008){Martin}, {de Jong}, \& {Rix}}]{Martin08}
{Martin}, N.~F., {de Jong}, J. T.~A., \& {Rix}, H.-W. 2008, \apj, 684, 1075

\bibitem[{{Martin} {et~al.}(2013{\natexlab{a}}){Martin}, {Ibata}, {McConnachie}, {Mackey}, {Ferguson}, {Irwin}, {Lewis}, \& {Fardal}}]{Martin2013}
{Martin}, N.~F., {Ibata}, R.~A., {McConnachie}, A.~W., {et~al.} 2013{\natexlab{a}}, \apj, 776, 80

\bibitem[{{Martin} {et~al.}(2009){Martin}, {McConnachie}, {Irwin}, {Widrow}, {Ferguson}, {Ibata}, {Dubinski}, {Babul}, {Chapman}, {Fardal}, {Lewis}, {Navarro}, \& {Rich}}]{Martin2009}
{Martin}, N.~F., {McConnachie}, A.~W., {Irwin}, M., {et~al.} 2009, \apj, 705, 758

\bibitem[{{Martin} {et~al.}(2013{\natexlab{b}}){Martin}, {Schlafly}, {Slater}, {Bernard}, {Rix}, {Bell}, {Ferguson}, {Finkbeiner}, {Laevens}, {Burgett}, {Chambers}, {Draper}, {Hodapp}, {Kaiser}, {Kudritzki}, {Magnier}, {Metcalfe}, {Morgan}, {Price}, {Tonry}, {Wainscoat}, \& {Waters}}]{Martin2013a}
{Martin}, N.~F., {Schlafly}, E.~F., {Slater}, C.~T., {et~al.} 2013{\natexlab{b}}, \apjl, 779, L10

\bibitem[{{Martin} {et~al.}(2013{\natexlab{c}}){Martin}, {Slater}, {Schlafly}, {Morganson}, {Rix}, {Bell}, {Laevens}, {Bernard}, {Ferguson}, {Finkbeiner}, {Burgett}, {Chambers}, {Hodapp}, {Kaiser}, {Kudritzki}, {Magnier}, {Morgan}, {Price}, {Tonry}, \& {Wainscoat}}]{Martin2013b}
{Martin}, N.~F., {Slater}, C.~T., {Schlafly}, E.~F., {et~al.} 2013{\natexlab{c}}, \apj, 772, 15

\bibitem[{{Mart{\'\i}nez-Delgado} {et~al.}(2022){Mart{\'\i}nez-Delgado}, {Karim}, {Charles}, {Boschin}, {Monelli}, {Collins}, {Donatiello}, \& {Alfaro}}]{Martinez-Delgado2022}
{Mart{\'\i}nez-Delgado}, D., {Karim}, N., {Charles}, E. J.~E., {et~al.} 2022, \mnras, 509, 16

\bibitem[{{Mart{\'\i}nez-Delgado} {et~al.}(2021){Mart{\'\i}nez-Delgado}, {Makarov}, {Javanmardi}, {Pawlowski}, {Makarova}, {Donatiello}, {Lang}, {Rom{\'a}n}, {Vivas}, \& {Carballo-Bello}}]{Martinez-Delgado21}
{Mart{\'\i}nez-Delgado}, D., {Makarov}, D., {Javanmardi}, B., {et~al.} 2021, \aap, 652, A48

\bibitem[{{McClure-Griffiths} {et~al.}(2009){McClure-Griffiths}, {Pisano}, {Calabretta}, {Ford}, {Lockman}, {Staveley-Smith}, {Kalberla}, {Bailin}, {Dedes}, {Janowiecki}, {Gibson}, {Murphy}, {Nakanishi}, \& {Newton-McGee}}]{McClure-Griffiths09}
{McClure-Griffiths}, N.~M., {Pisano}, D.~J., {Calabretta}, M.~R., {et~al.} 2009, \apjs, 181, 398

\bibitem[{{McConnachie}(2012)}]{McConnachie2012}
{McConnachie}, A.~W. 2012, \aj, 144, 4

\bibitem[{{McConnachie} {et~al.}(2018){McConnachie}, {Ibata}, {Martin}, {Ferguson}, {Collins}, {Gwyn}, {Irwin}, {Lewis}, {Mackey}, {Davidge}, {Arias}, {Conn}, {C{\^o}t{\'e}}, {Crnojevic}, {Huxor}, {Penarrubia}, {Spengler}, {Tanvir}, {Valls-Gabaud}, {Babul}, {Barmby}, {Bate}, {Bernard}, {Chapman}, {Dotter}, {Harris}, {McMonigal}, {Navarro}, {Puzia}, {Rich}, {Thomas}, \& {Widrow}}]{McConnachie2018}
{McConnachie}, A.~W., {Ibata}, R., {Martin}, N., {et~al.} 2018, \apj, 868, 55

\bibitem[{{McLeod} {et~al.}(2015){McLeod}, {Geary}, {Conroy}, {Fabricant}, {Ordway}, {Szentgyorgyi}, {Amato}, {Ashby}, {Caldwell}, {Curley}, {Gauron}, {Holman}, {Norton}, {Pieri}, {Roll}, {Weaver}, {Zajac}, {Palunas}, \& {Osip}}]{McLeod15}
{McLeod}, B., {Geary}, J., {Conroy}, M., {et~al.} 2015, \pasp, 127, 366

\bibitem[{{McNanna} {et~al.}(2023){McNanna}, {Bechtol}, {Mau}, {Nadler}, {Medoff}, {Drlica-Wagner}, {Cerny}, {Crnojevic}, {Mutlu-Pakdil}, {Vivas}, {Pace}, {Carlin}, {Collins}, {Ferguson}, {Martinez-Delgado}, {Martinez-Vazquez}, {Noel}, {Riley}, {Sand}, {Smercina}, {Tollerud}, {Wechsler}, {Abbott}, {Aguena}, {Alves}, {Bacon}, {Bom}, {Brooks}, {Burke}, {Carballo-Bello}, {Carnero Rosell}, {Carretero}, {da Costa}, {Davis}, {De Vicente}, {Diehl}, {Doel}, {Ferrero}, {Frieman}, {Giannini}, {Gruen}, {Gutierrez}, {Gruendl}, {Hinton}, {Hollowood}, {Honscheid}, {James}, {Kuehn}, {Marshall}, {Mena-Fernandez}, {Miquel}, {Pereira}, {Pieres}, {Plazas Malagon}, {Sakowska}, {Sanchez}, {Sanchez Cid}, {Santiago}, {Sevilla-Noarbe}, {Smith}, {Stringfellow}, {Suchyta}, {Swanson}, {Tarle}, {Weaverdyck}, \& {Wiseman}}]{McNanna2023}
{McNanna}, M., {Bechtol}, K., {Mau}, S., {et~al.} 2023, arXiv e-prints, arXiv:2309.04467

\bibitem[{{McQuinn} {et~al.}(2015){McQuinn}, {Skillman}, {Dolphin}, {Cannon}, {Salzer}, {Rhode}, {Adams}, {Berg}, {Giovanelli}, {Girardi}, \& {Haynes}}]{McQuinn2015}
{McQuinn}, K. B.~W., {Skillman}, E.~D., {Dolphin}, A., {et~al.} 2015, \apj, 812, 158

\bibitem[{{Mu{\~n}oz} {et~al.}(2018){Mu{\~n}oz}, {C{\^o}t{\'e}}, {Santana}, {Geha}, {Simon}, {Oyarz{\'u}n}, {Stetson}, \& {Djorgovski}}]{Munoz18}
{Mu{\~n}oz}, R.~R., {C{\^o}t{\'e}}, P., {Santana}, F.~A., {et~al.} 2018, \apj, 860, 66

\bibitem[{{M{\"u}ller} \& {Crosby}(2023)}]{Muller2023}
{M{\"u}ller}, O., \& {Crosby}, E. 2023, arXiv e-prints, arXiv:2307.09015

\bibitem[{{M{\"u}ller} {et~al.}(2015){M{\"u}ller}, {Jerjen}, \& {Binggeli}}]{Muller2015}
{M{\"u}ller}, O., {Jerjen}, H., \& {Binggeli}, B. 2015, \aap, 583, A79

\bibitem[{{M{\"u}ller} {et~al.}(2017){M{\"u}ller}, {Jerjen}, \& {Binggeli}}]{Muller2017}
---. 2017, \aap, 597, A7

\bibitem[{{M{\"u}ller} {et~al.}(2019){M{\"u}ller}, {Rejkuba}, {Pawlowski}, {Ibata}, {Lelli}, {Hilker}, \& {Jerjen}}]{Muller2019}
{M{\"u}ller}, O., {Rejkuba}, M., {Pawlowski}, M.~S., {et~al.} 2019, \aap, 629, A18

\bibitem[{{Munshi} {et~al.}(2019){Munshi}, {Brooks}, {Christensen}, {Applebaum}, {Holley-Bockelmann}, {Quinn}, \& {Wadsley}}]{Munshi19}
{Munshi}, F., {Brooks}, A.~M., {Christensen}, C., {et~al.} 2019, \apj, 874, 40

\bibitem[{{Mutlu-Pakdil} {et~al.}(2018){Mutlu-Pakdil}, {Sand}, {Carlin}, {Spekkens}, {Caldwell}, {Crnojevi{\'c}}, {Hughes}, {Willman}, \& {Zaritsky}}]{MutluPakdil2018}
{Mutlu-Pakdil}, B., {Sand}, D.~J., {Carlin}, J.~L., {et~al.} 2018, \apj, 863, 25

\bibitem[{{Mutlu-Pakdil} {et~al.}(2021){Mutlu-Pakdil}, {Sand}, {Crnojevi{\'c}}, {Drlica-Wagner}, {Caldwell}, {Guhathakurta}, {Seth}, {Simon}, {Strader}, \& {Toloba}}]{Mutlupakdil21}
{Mutlu-Pakdil}, B., {Sand}, D.~J., {Crnojevi{\'c}}, D., {et~al.} 2021, arXiv e-prints, arXiv:2105.01658

\bibitem[{{Mutlu-Pakdil} {et~al.}(2022){Mutlu-Pakdil}, {Sand}, {Crnojevi{\'c}}, {Jones}, {Caldwell}, {Guhathakurta}, {Seth}, {Simon}, {Spekkens}, {Strader}, \& {Toloba}}]{MutluPakdil22}
---. 2022, \apj, 926, 77

\bibitem[{{Nadler} {et~al.}(2021){Nadler}, {Drlica-Wagner}, {Bechtol}, {Mau}, {Wechsler}, {Gluscevic}, {Boddy}, {Pace}, {Li}, {McNanna}, {Riley}, {Garc{\'\i}a-Bellido}, {Mao}, {Green}, {Burke}, {Peter}, {Jain}, {Abbott}, {Aguena}, {Allam}, {Annis}, {Avila}, {Brooks}, {Carrasco Kind}, {Carretero}, {Costanzi}, {da Costa}, {De Vicente}, {Desai}, {Diehl}, {Doel}, {Everett}, {Evrard}, {Flaugher}, {Frieman}, {Gerdes}, {Gruen}, {Gruendl}, {Gschwend}, {Gutierrez}, {Hinton}, {Honscheid}, {Huterer}, {James}, {Krause}, {Kuehn}, {Kuropatkin}, {Lahav}, {Maia}, {Marshall}, {Menanteau}, {Miquel}, {Palmese}, {Paz-Chinch{\'o}n}, {Plazas}, {Romer}, {Sanchez}, {Scarpine}, {Serrano}, {Sevilla-Noarbe}, {Smith}, {Soares-Santos}, {Suchyta}, {Swanson}, {Tarle}, {Tucker}, {Walker}, {Wester}, \& {DES Collaboration}}]{Nadler2021}
{Nadler}, E.~O., {Drlica-Wagner}, A., {Bechtol}, K., {et~al.} 2021, \prl, 126, 091101

\bibitem[{{Okamoto} {et~al.}(2019){Okamoto}, {Arimoto}, {Ferguson}, {Irwin}, {Bernard}, \& {Utsumi}}]{Okamoto2019}
{Okamoto}, S., {Arimoto}, N., {Ferguson}, A. M.~N., {et~al.} 2019, \apj, 884, 128

\bibitem[{{Papastergis} {et~al.}(2015){Papastergis}, {Giovanelli}, {Haynes}, \& {Shankar}}]{Papastergis2015}
{Papastergis}, E., {Giovanelli}, R., {Haynes}, M.~P., \& {Shankar}, F. 2015, \aap, 574, A113

\bibitem[{{Park} {et~al.}(2019){Park}, {Moon}, {Zaritsky}, {Kim}, {Lee}, {Cha}, \& {Lee}}]{Park2019}
{Park}, H.~S., {Moon}, D.-S., {Zaritsky}, D., {et~al.} 2019, \apj, 885, 88

\bibitem[{{Park} {et~al.}(2017){Park}, {Moon}, {Zaritsky}, {Pak}, {Lee}, {Kim}, {Kim}, \& {Cha}}]{Park2017}
---. 2017, \apj, 848, 19

\bibitem[{{Radburn-Smith} {et~al.}(2011){Radburn-Smith}, {de Jong}, {Seth}, {Bailin}, {Bell}, {Brown}, {Bullock}, {Courteau}, {Dalcanton}, {Ferguson}, {Goudfrooij}, {Holfeltz}, {Holwerda}, {Purcell}, {Sick}, {Streich}, {Vlajic}, \& {Zucker}}]{Radburn-Smith2011}
{Radburn-Smith}, D.~J., {de Jong}, R.~S., {Seth}, A.~C., {et~al.} 2011, \apjs, 195, 18

\bibitem[{{Rizzi} {et~al.}(2007){Rizzi}, {Tully}, {Makarov}, {Makarova}, {Dolphin}, {Sakai}, \& {Shaya}}]{Rizzi2007}
{Rizzi}, L., {Tully}, R.~B., {Makarov}, D., {et~al.} 2007, \apj, 661, 815

\bibitem[{{Rockosi} {et~al.}(2002){Rockosi}, {Odenkirchen}, {Grebel}, {Dehnen}, {Cudworth}, {Gunn}, {York}, {Brinkmann}, {Hennessy}, \& {Ivezi{\'c}}}]{Rockosi2002}
{Rockosi}, C.~M., {Odenkirchen}, M., {Grebel}, E.~K., {et~al.} 2002, \aj, 124, 349

\bibitem[{{Salaris} {et~al.}(2002){Salaris}, {Cassisi}, \& {Weiss}}]{Salaris2002}
{Salaris}, M., {Cassisi}, S., \& {Weiss}, A. 2002, \pasp, 114, 375

\bibitem[{{Sales} {et~al.}(2022){Sales}, {Wetzel}, \& {Fattahi}}]{Sales2022}
{Sales}, L.~V., {Wetzel}, A., \& {Fattahi}, A. 2022, Nature Astronomy, 6, 897

\bibitem[{{Sand} {et~al.}(2015){Sand}, {Spekkens}, {Crnojevi{\'c}}, {Hargis}, {Willman}, {Strader}, \& {Grillmair}}]{Sand15b}
{Sand}, D.~J., {Spekkens}, K., {Crnojevi{\'c}}, D., {et~al.} 2015, \apjl, 812, L13

\bibitem[{{Sand} {et~al.}(2014){Sand}, {Crnojevi{\'c}}, {Strader}, {Toloba}, {Simon}, {Caldwell}, {Guhathakurta}, {McLeod}, \& {Seth}}]{Sand14}
{Sand}, D.~J., {Crnojevi{\'c}}, D., {Strader}, J., {et~al.} 2014, \apjl, 793, L7

\bibitem[{{Savino} {et~al.}(2022){Savino}, {Weisz}, {Skillman}, {Dolphin}, {Kallivayalil}, {Wetzel}, {Anderson}, {Besla}, {Boylan-Kolchin}, {Bullock}, {Cole}, {Collins}, {Cooper}, {Deason}, {Dotter}, {Fardal}, {Ferguson}, {Fritz}, {Geha}, {Gilbert}, {Guhathakurta}, {Ibata}, {Irwin}, {Jeon}, {Kirby}, {Lewis}, {Mackey}, {Majewski}, {Martin}, {McConnachie}, {Patel}, {Rich}, {Simon}, {Sohn}, {Tollerud}, \& {van der Marel}}]{Savino2022}
{Savino}, A., {Weisz}, D.~R., {Skillman}, E.~D., {et~al.} 2022, \apj, 938, 101

\bibitem[{{Schlafly} \& {Finkbeiner}(2011)}]{Schlafly11}
{Schlafly}, E.~F., \& {Finkbeiner}, D.~P. 2011, \apj, 737, 103

\bibitem[{{Schlegel} {et~al.}(1998){Schlegel}, {Finkbeiner}, \& {Davis}}]{Schlegel98}
{Schlegel}, D.~J., {Finkbeiner}, D.~P., \& {Davis}, M. 1998, \apj, 500, 525

\bibitem[{{Sharina} {et~al.}(2008){Sharina}, {Karachentsev}, {Dolphin}, {Karachentseva}, {Tully}, {Karataeva}, {Makarov}, {Makarova}, {Sakai}, {Shaya}, {Nikolaev}, \& {Kuznetsov}}]{Sharina2008}
{Sharina}, M.~E., {Karachentsev}, I.~D., {Dolphin}, A.~E., {et~al.} 2008, \mnras, 384, 1544

\bibitem[{{Smercina} {et~al.}(2018){Smercina}, {Bell}, {Price}, {D'Souza}, {Slater}, {Bailin}, {Monachesi}, \& {Nidever}}]{Smercina18}
{Smercina}, A., {Bell}, E.~F., {Price}, P.~A., {et~al.} 2018, \apj, 863, 152

\bibitem[{{Smercina} {et~al.}(2022){Smercina}, {Bell}, {Samuel}, \& {D'Souza}}]{Smercina2022}
{Smercina}, A., {Bell}, E.~F., {Samuel}, J., \& {D'Souza}, R. 2022, \apj, 930, 69

\bibitem[{{Springob} {et~al.}(2005){Springob}, {Haynes}, {Giovanelli}, \& {Kent}}]{Springob2005}
{Springob}, C.~M., {Haynes}, M.~P., {Giovanelli}, R., \& {Kent}, B.~R. 2005, \apjs, 160, 149

\bibitem[{{Stetson}(1987)}]{Stetson87}
{Stetson}, P.~B. 1987, \pasp, 99, 191

\bibitem[{{Stetson}(1994)}]{Stetson94}
---. 1994, \pasp, 106, 250

\bibitem[{{Toloba} {et~al.}(2016){Toloba}, {Sand}, {Spekkens}, {Crnojevi{\'c}}, {Simon}, {Guhathakurta}, {Strader}, {Caldwell}, {McLeod}, \& {Seth}}]{Toloba16}
{Toloba}, E., {Sand}, D.~J., {Spekkens}, K., {et~al.} 2016, \apjl, 816, L5

\bibitem[{{Tully} {et~al.}(2006){Tully}, {Rizzi}, {Dolphin}, {Karachentsev}, {Karachentseva}, {Makarov}, {Makarova}, {Sakai}, \& {Shaya}}]{Tully2006}
{Tully}, R.~B., {Rizzi}, L., {Dolphin}, A.~E., {et~al.} 2006, \aj, 132, 729

\bibitem[{{Walsh} {et~al.}(2009){Walsh}, {Willman}, \& {Jerjen}}]{Walsh2009}
{Walsh}, S.~M., {Willman}, B., \& {Jerjen}, H. 2009, \aj, 137, 450

\bibitem[{{Wechsler} \& {Tinker}(2018)}]{Wechsler2018}
{Wechsler}, R.~H., \& {Tinker}, J.~L. 2018, \araa, 56, 435

\bibitem[{{Weisz} {et~al.}(2011){Weisz}, {Dalcanton}, {Williams}, {Gilbert}, {Skillman}, {Seth}, {Dolphin}, {McQuinn}, {Gogarten}, {Holtzman}, {Rosema}, {Cole}, {Karachentsev}, \& {Zaritsky}}]{Weisz2011}
{Weisz}, D.~R., {Dalcanton}, J.~J., {Williams}, B.~F., {et~al.} 2011, \apj, 739, 5

\bibitem[{{Westmeier} {et~al.}(2017){Westmeier}, {Obreschkow}, {Calabretta}, {Jurek}, {Koribalski}, {Meyer}, {Musaeva}, {Popping}, {Staveley-Smith}, {Wong}, \& {Wright}}]{Westmeier2017}
{Westmeier}, T., {Obreschkow}, D., {Calabretta}, M., {et~al.} 2017, \mnras, 472, 4832

\bibitem[{{Yaryura} {et~al.}(2016){Yaryura}, {Helmi}, {Abadi}, \& {Starkenburg}}]{Yaryura2016}
{Yaryura}, C.~Y., {Helmi}, A., {Abadi}, M.~G., \& {Starkenburg}, E. 2016, \mnras, 457, 2415

\end{thebibliography}

\end{document}